\pdfoutput=1  
\documentclass[
  aps,%
  prx,
 twocolumn,%
 groupedaddress,%
 superscriptaddress,%
  showpacs,%
 letterpaper,%
 amsfonts,%
 footinbib,%
  10pt,%
 floatfix,%
]{revtex4-2}
 
\usepackage{dcolumn} 
\usepackage{tikz}
\usetikzlibrary{positioning,arrows.meta}

\usepackage{hyperref}
\hypersetup{
    colorlinks=true,
    citecolor=blue,
    linkcolor=blue , 
    filecolor=cyan,      
    urlcolor=magenta,
    pdfcreator = {\LaTeX\ and \flqq hyperref\frqq},
}

\RequirePackage{graphicx,amsmath,amssymb,bm}
\usepackage{bbm}
\usepackage{float}
\usepackage{framed} 
\usepackage{amsthm}
\usepackage{caption}
\usepackage{subcaption}
\usepackage[normalem]{ulem}
\usepackage{comment}
\usepackage{slashed} 
\usepackage{upgreek}
\usepackage{xcolor}
\usepackage{enumitem}
\usepackage{tikz-3dplot}  
\usepackage{yhmath}

\graphicspath{ {./Figs/} }

\newcommand{\bb}{\mathbb}
\newcommand{\tr}{\text{tr}}

 \newcommand{\e}{\text{e}}

\newcommand{\Pp}{{\cal{P}}}

\newcommand{\bbc }{\bb{C}}

\newcommand{\bbz}{\bb{Z}}

\def\be{\begin{equation}}
\def\ee{\end{equation}} 
\def\bsh{\begin{shaded}}
\def\esh{\end{shaded}} 
\def\bpm{\begin{pmatrix}}
\def\epm{\end{pmatrix}}

\begin{document} 

\title{Exact fermionic dual of the Bose-Hubbard model}
\author{Lei Su}
\affiliation{Division of Condensed Matter Physics and Materials Science, Brookhaven National Laboratory, Upton, New York 11973, USA}
\affiliation{Department of Physics, University of Chicago, Chicago, Illinois 60637, USA}
\affiliation{Pritzker School of Molecular Engineering, University of Chicago, Chicago, Illinois 60637, USA}
\author{Ivar Martin}
\affiliation{Department of Physics, University of Chicago, Chicago, Illinois 60637, USA}

\affiliation{Materials Science Division, Argonne National Laboratory, Lemont, Illinois 60439, USA}
\author{Aashish A. Clerk}
\affiliation{Pritzker School of Molecular Engineering, University of Chicago, Chicago, Illinois 60637, USA}

\begin{abstract}
Recent developments have established exact bosonization and fermionization with a $\mathbb{Z}_2$ symmetry as dualities through gauging. In this work, we apply fermionic gauging, which realizes generalized Jordan-Wigner transformations, to the Bose-Hubbard (BH) model with a global $U(1)$ symmetry and derive an exact dual description in terms of fermionic composites, built from bosons and fermions. In 1D, this duality generalizes the exact mapping between the extended hard-core BH model and the spinless Fermi-Hubbard model to include soft-core bosons. At low energies, the mapping reduces to the well-known equivalence between the sine-Gordon model and the Thirring model. The oscillation wave vector of the fermionic composite correlation function in the gapless phase is fixed by their density, providing a novel manifestation of Luttinger's theorem. We verify the exact duality using density matrix renormalization group  (DMRG) calculations and  demonstrate that the gapless phase and the phase transition are governed by the compact boson conformal field theory. Our construction naturally extends  to generic bosonic systems and higher dimensions, opening new avenues for studying Bose-Fermi mixtures in optical lattices and other strongly correlated quantum systems.
\end{abstract} 
\maketitle

\tableofcontents

\section{Introduction}
Dualities relate physical theories that may appear different yet describe the same underlying phenomena. A classic example is  electromagnetic duality in  Maxwell's equations, which interchanges the electric and magnetic sectors, relating charges and monopoles. Generalizations of electromagnetic duality, commonly referred to as ``S-duality," appear in a wide range of physical theories \cite{figueroa1998electromagnetic} and in mathematical frameworks \cite{kapustinElectricmagnetic2007}. Another prominent example in $(1+1)$D is the duality between the massive Thirring model and the sine-Gordon model \cite{coleman1975}, which can be understood as a consequence of bosonization and fermionization \cite{gogolin2004bosonization, giamarchiQuantum2003}. At the lattice level, this correspondence emerges as the low-energy limit of the exact equivalence between the extended hard-core Bose-Hubbard (BH) model and the extended spinless Fermi-Hubbard model on the lattice \cite{cazalillaOne2011a}.

Exact bosonization and fermionization on 1D chains realized through the Jordan-Wigner (JW) transformation \cite{jordan1928} have been known for nearly a century. A modern perspective interprets it as gauging a $\bbz_2$ symmetry of the system, including the $\bbz_2^F$ fermion parity symmetry in fermionic systems. This $\bbz_2$-gauging framework has since been generalized to higher dimensions \cite{suBosonization2026, su$mathbbZ_2$2025a}. In particular, fermionization via the JW transformation can be viewed as a fermionic gauging procedure, in which a pair of Majorana fermions replaces a conventional Ising spin as the ``gauge field," followed by a unitary transformation that decouples the original bosonic degrees of freedom.  Thus far,  exact fermionization through $\bbz_2$-fermionic gauging has been explored primarily for systems with a $\bbz_2$ symmetry. In this work, we extend this framework to bosonic systems with a $U(1)$ symmetry.

A simple example of a bosonic model with a global $U(1)$ symmetry is the BH model \cite{cazalillaOne2011a}, which can be realized in condensed matter systems and ultracold atoms in optical lattices. At commensurate fillings, the BH may undergo a quantum phase transition from a superfluid phase to an insulating phase as the onsite interaction strength is increased. Gauging the $\bbz_2$ subgroup of the $U(1)$ symmetry in the conventional way using Ising spins was recently shown to produce exotic phases and phase transitions \cite{su2024}.  We show in this work that the less-conventional fermionic gauging procedure reveals new surprises, in particular an exact fermionic dual of the BH model described in terms of boson-fermion composite operators. In 1D, such a transformation extends the exact boson-fermion duality of the hard-core limit to the full BH model. Fermionic gauging effectively projects out the bosonic levels with odd occupancy and replaces them with fermionic levels. Since the local bosonic Hilbert space is infinite-dimensional, whereas the local fermionic Hilbert space is two-dimensional, the fermionic character of the system is most pronounced at low fillings.

The duality we establish can also be understood as a generalized JW transformation. Fermionic gauging provides a systematic decomposition of bosonic operators dressed by a JW string into boson-fermion composite operators, connecting our construction to the broader context of Bose-Fermi mixtures in optical lattices and other strongly correlated systems \cite{Lewenstein2004, Gunter2006, Pollet2006, Sengupta2007, sugawaInteraction2011, baroniQuantum2024}. The basic idea is easy to state: we can approximate the string-dressed bosonic annihilation operator $b_j$ by the composite operator $\sqrt{2}\tilde{b}_{j} f^{\dagger}_{j} +  f_{j}$ for $\langle \tilde{n}_{b, j}\rangle \ll 1$, where $\tilde{b}_j^{\dagger}$ is a charge-2 canonical boson operator and $f_j^{\dagger}$ is a charge-1 canonical fermion operator. The composite operator has a unit charge and is fermionic in nature.
Under this mapping, nearest-neighbor bosonic hopping terms will generate charge-conserving pairing terms $\sim \tilde{b}^{\dagger}_{j} f_{j} f_{j+1} + h.c.$.  Our duality thus reveals a surprising connection between the BH model and Bose-Fermi models of charge-conserving pairing and atom-molecule conversion. The boson-fermion composite operators discussed in this work can potentially be generalized to boson-anyon composites by replacing the JW string with a fractional version, leading to the anyon-Hubbard model in 1D, which has been realized in experiments in recent years
\cite{kunduExact1999, keilmannStatistically2011, Greschner2015, strater2016, zhangGroundstate2017, kwanRealization2024, dharObserving2025, bakkali-hassaniRevealing2026}. 

Our paper is organized as follows. In Sec.~\ref{sec:fermionicdual}, we derive the fermionic dual of the 1D BH model using  fermionic gauging and elucidate its connection to the generalized JW transformation, thereby extending the known mapping between the extended hard-core BH model and the spinless Fermi-Hubbard model with nearest-neighbor interactions. 
In Sec.~\ref{sec:eft}, we discuss Luttinger liquid theory and the compact boson conformal field theory (CFT), and argue that a canonical fermionic field emerges from noncanonical fermionic composites in the low-energy theory in the gapless phase, generalizing the equivalence between the sine-Gordon model and the Thirring model. We also identify several microscopic operators with their corresponding continuum fields, and argue that a generalized Luttinger's theorem applies to the fermionic composites. In Sec.~\ref{sec:numerics}, we present density matrix renormalization group (DMRG) calculations to confirm the spectral equivalence between the dual theories and verify these operator identifications through correlation functions in the gapless phase. We focus on two representative cases, half filling and unit filling, and discuss the subtlety associated with edge-to-edge correlations. In Sec.~\ref{sec:2dgen}, we present the 2D generalization of the fermionic dual on the square and triangular lattices and discuss its connection to Majorana fermion surface codes. In Sec.~\ref{sec:dicussion}, we discuss our main results and conclude. Some technical details are relegated to the Appendices. The fermionic gauging procedure for a $\bbz_2$ subgroup naturally generalizes to other models, with representative examples presented in the Appendices.  For completeness, we also discuss the corresponding bosonic gauging constructions in the Appendices.

\section{Fermionic dual of 1D Bose-Hubbard model}
\label{sec:fermionicdual}
The Hamiltonian of the 1D BH model is given by 
\be
   H =  -t \sum_j ( b^{\dagger}_j b_{j+1} +h.c.)  + \frac{U}{2} \sum_j n_{b, j}  (n_{b, j} -1),  
   \label{eq:BH}
\ee 
where $b_j^{\dagger}$ ($b_j$) creates (annihilates) a canonical boson with onsite occupation number $n_{b,j} = b_j^{\dagger} b_j$. Here, $t$ denotes the hopping amplitude and $U$ the onsite repulsive interaction. The model has a $U(1)$ symmetry associated with the particle-number conservation. Although our analysis can be presented in a more general setting, we consider a fixed number of particles. At incommensurate fillings, the ground state remains in a superfluid phase for finite $U/t$. At certain commensurate fillings, however,
it may undergo a Berezinskii-Kosterlitz-Thouless (BKT) transition \cite{berezinskii1971destruction, Kosterlitz1973}  from  a superfluid phase at small $U/t$ to a Mott insulating phase at large $U/t$. At unit filling, for example, quantum phase slips proliferate at the transition and destroy the superfluid order. At half filling, in contrast, such phase slips are not sufficiently relevant to drive a transition, and the system remains in a gapless phase \cite{cazalillaOne2011a}.

\begin{figure}[tb]
    \centering
    \includegraphics[width=0.7\linewidth]{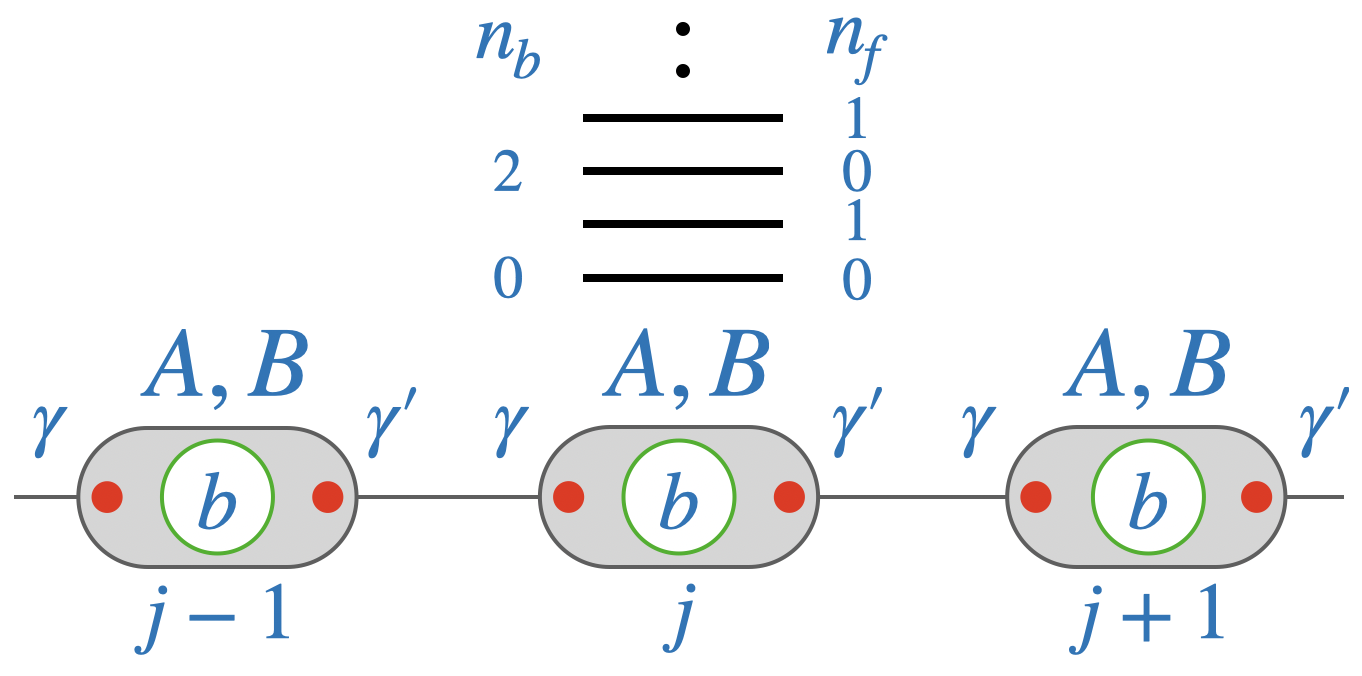}
    \caption{Fermionic gauging of the 1D Bose-Hubbard model. One $b$ boson resides on each site. A pair of Majorana fermions, $\gamma'_{j}$ and $\gamma_{j+1}$, is inserted between two neighboring bosons, and a Gauss law is imposed at each site. The gauging projects out the onsite sub-Hilbert space with odd boson number $n_b$ and replaces it with a $\bbz_2$-graded fermionic level $n_f$, yielding a new fermionic composite $A$ or $B$.}
    \label{fig1}
\end{figure}

\subsection{Fermionic gauging}
To gauge a $\bbz_2$ subgroup of $U(1)$, one typically introduces Ising gauge fields and imposes a Gauss law on the minimally coupled Hamiltonian \cite{su2024}. In principle, a disentangling unitary transformation can be applied to decouple the matter and gauge degrees of freedom in the Gauss law, thereby simplifying the resulting model. We discuss this procedure in Appendix \ref{sec:bosonic_gauging}. In this work, we instead insert a pair of Majorana fermions, denoted by $\gamma'_{j}$ and $\gamma_{j+1}$, as the gauge degrees of freedom on each edge (shown in Fig.~\ref{fig1}), and impose the following Gauss law on each site $G_j = (-1)^{n_{b, j}} (-1)^{n_{f,j}} =1$.
Here,  $(-1)^{n_{f,j}} = i\gamma_{j} \gamma'_{j}$  with $ \gamma_{j} = (f_{j} + f_{j}^{\dagger})$ and $ \gamma'_{j} = i(f_j - f_j^{\dagger})$.   The minimally coupled Hamiltonian is 
\be 
 H'  =  -t \sum_j (b^{\dagger}_j \Xi_{j+1/2} b_{j+1} + h.c.)   + \frac{U}{2} \sum_j n_{b, j}  (n_{b, j} -1) 
\ee  
with $\Xi_{j+1/2} \equiv i \gamma'_{j} \gamma_{j+1} $. One can check that $H'$ commutes with the Gauss law. 

We can now perform a unitary disentangling transformation 
\be 
\tilde{U} = \prod_j(\Pp_{j}^+ +\Pp_{j}^- K_j),
\ee 
where $\Pp_{j}^{\pm} = [1\pm (-1)^{n_{f,j}}]/2$ is a fermion parity projection and $K= \sum_{k \in \bbz_{\ge 0}} |2k\rangle\langle 2k+1| +|2k+1\rangle\langle 2k|$ is an onsite unitary boson parity-switching operator, written in terms of bosonic Fock states (see Appendix~\ref{sec:decomposition}). The Gauss law becomes 
$(-1)^{n_{b, j}} =1$, which restricts the onsite boson number $n_{b, j}$  to even integers. Also, within the even sector, $ \tilde{U} n_{b, j} \tilde{U}^{-1}= n_{b, j} + n_{f, j}$, which generates the new $U(1)$ symmetry that includes the fermion parity generated by $\prod_j (-1)^{n_{f, j}}$ as a $\bbz_2$ subgroup. Under $\tilde{U}$,   
\be 
\begin{split}
& b^{\dagger}_j \Xi_{j+1/2} b_{j+1} \to  -[b_j^{\dagger 2}  \frac{1}{\sqrt{n_{b, j} +1}} f_{j} - \sqrt{n_{b, j} +1} f^{\dagger}_{j} ]  \\
& \times [\frac{1}{\sqrt{n_{b, j+1} +1}}  b_{j+1}^2 f^{\dagger}_{j+1} + \sqrt{n_{b, j+1} +1}  f_{j+1}].  
\end{split}
\ee 
Here, $f(-1)^{n_f} = - f$ and $f^{\dagger} (-1)^{n_f}= f^{\dagger}$ are used. 

We can simplify the above expression by introducing new canonical boson operators  
$ \tilde{b}_j = [2(n_{b,j}+1)]^{-1/2} b_j^{2}, \quad  \tilde{b}_j^{\dagger} =  b^{\dagger 2}_j  [2(n_{b, j}+1)]^{-1/2}$.  Then  $  
\tilde{n}_{b, j} = \tilde{b}_j^{\dagger} \tilde{b}_j =   n_{b, j}/2$ and $ [\tilde{b}_j , \tilde{b}_j^{\dagger} ] = 1$ within the even sector.  Next, we use these new boson operators to introduce new composite operators defined as
\be 
A_j = \sqrt{2}\tilde{b}_{j} f^{\dagger}_{j} + \sqrt{2 \tilde{n}_{b, j} +1} f_{j}
\ee  and
\be 
B_j = \sqrt{2}\tilde{b}_{j} f^{\dagger}_{j} - \sqrt{2 \tilde{n}_{b, j} +1} f_{j} =A_j (-1)^{n_{f, j}}.
\ee 
To gain intuition into these operators, consider their action in Fock space.  
We define $|\psi_{2n+s}\rangle \equiv |n, s\rangle$, where $n$ is the onsite $\tilde{b}$-boson number and $s = 0, 1$ is the fermion number, as shown in Fig.~\ref{fig1}. Then $A_j|\psi_{n, j}\rangle = \sqrt{n}|\psi_{n-1, j} \rangle$, $A_j^{\dagger}|\psi_{n, j}\rangle = \sqrt{n+1}|\psi_{n+1, j}\rangle $, and $\tilde{n}_j \equiv A_j^{\dagger} A_j =2\tilde{n}_{b, j} + n_{f, j}$. These operators satisfy the algebra $[A_j, A_j^{\dagger}] = 1$ and $ \{A_i, A_j\} =\{A_i ,  A_j^{\dagger}\} = 0$ for $ i \neq j$. The anticommutation relation between $A_i$ and $A_j^{\dagger}$ at different sites follows from the fact that $A_j$ is odd under fermion parity, which is also reflected in the onsite anticommutation relation $\{A_j, B_j\} =0$.  The novelty lies in the onsite commutation relation between $A_j$ and $A_j^{\dagger}$, which arises from its mixed nature.  Since $A_j^2 \neq 0$ and $A_j^{\dagger 2} \neq 0$, the fermionic composite $A_j$ is not a canonical fermion and does not obey the Pauli exclusion principle.  

With the new operators, the Hamiltonian can be written as  
\begin{align}
H''   =&  t  \sum_j    (B_j^{\dagger}A_{j+1}  +h.c.) 
   + \frac{U}{2} \sum_j \tilde{n}_j(\tilde{n}_j-1)
   \label{eq:dualBH} 
   \\
    = &  t  \sum_j  [(\sqrt{2}\tilde{b}^{\dagger}_{j} f_{j} - \sqrt{2 \tilde{n}_{b, j} +1} f_{j}^{\dagger}) \nonumber \\
   &\ \times  (\sqrt{2}\tilde{b}_{j+1} f^{\dagger}_{j+1} + \sqrt{2 \tilde{n}_{b, j+1} +1} f_{j+1}) + h.c.] \nonumber \\
&\  + \frac{U}{2} \sum_j (2\tilde{n}_{b, j} + n_{f, j})(2\tilde{n}_{b, j} + n_{f, j}-1). \nonumber
\end{align} 
Eq.~(\ref{eq:dualBH}) is a central result of this work: it describes the fermionic dual of the original BH model. Its relationship to the BH model is analogous to that between the Kitaev-Majorana chain and the 1D transverse-field Ising model. In this dual model of Bose-Fermi mixture, the fundamental charge-1 excitations are $f^{\dagger}$ and $f \tilde{b}^{\dagger}$ under the new $U(1)$ symmetry generated by $\sum_j \tilde{n}_j = \sum_j A_j^{\dagger} A_j =\sum_j2\tilde{n}_{b, j} + n_{f, j}$. Two $f$ fermions can be annihilated by creating a $\tilde{b}$ boson, and vice versa. The $f$ fermion sector and $\tilde{b}$ boson sector are thus connected. The composite structure of our dual model is reminiscent of (but distinct from) the slave-boson/fermion formalism \cite{leeDoping2006}. It arises because fermionic gauging is effectively  equivalent to stacking a gapped fermionic system onto the original bosonic model, followed by gauging the diagonal $\bbz_2 \subset U(1) \times \bbz_2^F$ \cite{su$mathbbZ_2$2025a}, thereby binding bosons and fermions together.

Further intuition into our model comes from the resemblance between its kinetic terms and the Hamiltonian of the Kitaev-Majorana chain, with the correspondence $\gamma_j \sim A_j$ and $\gamma'_j \sim B_j$. Indeed, in the gapless phase, if we take $\langle \tilde{b}_j \rangle \approx \alpha$ and $\langle \tilde{n}_{b, j} \rangle \approx |\alpha|^2$ at the mean-field level, $A_j$ and $B_j$ are linear combinations of fermionic particle and hole operators. If the filling factor of $\tilde{n}$ is large, $\langle \tilde{n}_{b, j} \rangle \approx |\alpha|^2 \gg 1$, and $A_j$ and $B_j$ reduce to (unnormalized) Majorana fermion operators after a phase rotation. Heuristically, $\tilde{b}_j$ may be interpreted as annihilating local Cooper pairs. 
While the above interpretation is appealing, in 1D, long-range fluctuations in $\langle \tilde{b}_j \rangle \approx \alpha$ severely limit its utility. We note that while our model has some features in common with the particle-number conserving model of topological superconductivity in Ref.~\cite{lapaRigorous2020}, the model in that work assumes the existence of a long-range order parameter $\alpha$.  In contrast, fluctuations in $\alpha$ are crucial to the physics of our model.    
Another difference from the Kitaev-Majorana model lies in boundary physics. Since $A_j^\dagger\neq A_j$ and $[B_j,A_j]\neq0$, the simple Majorana edge-mode construction does not carry over directly to our model. Numerically, we find no ground-state degeneracy for an open chain. We return to the boundary physics in later sections.

\subsection{Generalized Jordan-Wigner transformations}

To further emphasize the duality of the model in Eq.~(\ref{eq:dualBH}) to the BH model, we can map it back to the BH Hamiltonian in Eq.~(\ref{eq:BH}) using a more familiar approach. Define
\be b_j = \prod_{i <j}(-1)^{\tilde{n}_i} A_j.
\ee 
One can check that $b_j$ and $b_j^{\dagger}$ satisfy the canonical bosonic commutation relation and the Hamiltonian becomes simply the BH model. This map is a generalized JW transformation, with the inverse given by 
\be A_j = \prod_{i <j}(-1)^{n_{b, i}} b_j.
\label{eq:genJW}
\ee 
Note that we could have derived our dual model directly by applying this generalized JW transformation to the BH model.  However, this approach would obscure important aspects of the resulting model that are made explicit in the fermionic-gauging derivation. In particular, the latter provides explicit expressions for $A_j$ (or $B_j$) in terms of $\tilde{b}_j$ and $f_j$, which are not obtained from the JW derivation.

As an aside, a more generalized version of the JW transformation $\prod_{i <j}e^{i\theta n_{b, i}} b_j$, with arbitrary $\theta$,  maps bosons to anyons and has been discussed in Refs.~\cite{kunduExact1999, keilmannStatistically2011}. This transformation can be applied to the 1D anyon-Hubbard model, whose kinetic terms take the form $b^{\dagger}_j e^{i \theta n_{b, j}} b_{j+1} +h.c.$, featuring an additional occupation-dependent phase factor. For generic $\theta$, this phase generates many-body interactions even at $U =0$. The anyon-Hubbard model exhibits a variety of novel phases and has been actively studied both theoretically and experimentally \cite{keilmannStatistically2011, Greschner2015, strater2016, zhangGroundstate2017, kwanRealization2024, dharObserving2025, bakkali-hassaniRevealing2026}. In the special case  $\theta = \pi$, the kinetic term  $b^{\dagger}_j (-1)^{n_{b, j}} b_{j+1} +h.c.$, under a generalized JW transformation,  maps to   $A_j^{\dagger}A_{j+1}  +h.c.$, rather than $B_j^{\dagger}A_{j+1}  +h.c.$ as in Eq.~(\ref{eq:dualBH}).

The generalized JW transformation in Eq.~(\ref{eq:genJW}) reduces to the conventional one in the hard-core boson limit, where onsite occupancy $n_{b, j}$ is at most 1. Indeed, the hard-core constraint can be realized in the large onsite repulsion limit $U \to \infty$ in the BH model. Then $A_j$ reduces to $f_j$ and the model reduces to a free theory. To make it more interesting in this case, we can add nearest-neighbor interactions as in the extended BH model 
\be 
H = -t \sum_j ( b^{\dagger}_j b_{j+1} + b_j b_{j+1}^{\dagger})   + V \sum (n_j -\frac{1}{2}) (n_{j+1} -\frac{1}{2}).
\label{eq:extendedBH}
\ee 
It is well known that the above model can be mapped to the XXZ model and, in turn, to a spinless Fermi-Hubbard model via the conventional JW transformation. In the hard-core limit, the above fermionic gauging procedure reduces to this conventional mapping. The disentangling unitary maps $
b_j^{\dagger} f_j^{\dagger} \to   f_j^{\dagger} $, $ b_j f_j \to   f_j  $, $n_j \to n_{f,j}$,  and 
$b^{\dagger}_j \Xi_{j+1/2} b_{j+1}   \to f^{\dagger}_{j}  f_{j+1}$, thereby reproducing the spinless Fermi-Hubbard model. 

Moving away from the hard-core limit, at large but finite $U$, we can perform a large-$U$ expansion around the $U \to \infty$ limit. To leading order,  virtual bosonic fluctuations in Eq.~(\ref{eq:dualBH}) induce nearest-neighbor interactions, together with three-site interactions, among the $f$ fermions. Moreover, in the original Hilbert space, the operator that creates the low-energy quasiparticle is not the bare fermion, but rather a dressed fermionic operator involving nearby bosonic and fermionic degrees of freedom. See Appendix \ref{sec:largeU} for details.

One point deserves emphasis. At the algebraic level, the fermionic composite $A_j$ satisfies $[A_j, A_j^{\dagger}] = 1$, whereas the canonical fermion $f_j$ satisfies $\{f_j, f_j^{\dagger}\} =1$. Thus, the reduction $A_j \to f_j$ in the hard-core limit should be understood as a projection onto the low-energy sector. Projection onto the hard-core subspace does not commute with taking commutators. States created by $\tilde{b}_j^{\dagger} f_j$ acquire infinite energy in the hard-core limit and are suppressed dynamically, so they do not appear in low-energy effective time evolution. Equal-time commutators, however, still retain information about these states because they are exact operator identities and involve no energy denominators. In other words, the canonical fermion $f_j$ emerges from $A_j$ after projection onto the low-energy sector. As we argue below, this emergent description persists at finite $U$.

Alternatively, one can retrieve the BH model from Eq.~(\ref{eq:dualBH}) by gauging the (dual) fermion parity of the gauged BH model or applying the conventional JW  transformation to the fermions $f_j$ and $f_j^{\dagger}$: $\sigma_j^+ = \prod_{i <j} (-1)^{n_{f,i}} f_j^{\dagger} $ and $\sigma_j^- = \prod_{i <j} (-1)^{n_{f,i}} f_j$. Then 
\be 
A_j \to  b_j \equiv \sqrt{2}\tilde{b}_{j} \sigma^+_{j}  +  \sqrt{2 \tilde{n}_{b, j} +1} \sigma^-_{j}.
\label{eq:dep_b}
\ee  
One can verify that the new operator $b_j$ is a canonical bosonic annihilation operator, whose local Hilbert space $H_b$ decomposes as $H_{\tilde{b}} \otimes H_{\sigma}$, where the Pauli operators $\sigma^{\pm}_j$ connect sectors of $H_b$ with different bosonic parities. See also Appendix~\ref{sec:decomposition}. 

It is interesting to note that $H_b$ can be viewed as an extension of $H_{\tilde{b}}$ by a $\bbz_2$ bosonic parity degree of freedom, and this extension process can be iterated \footnote{This construction should be contrasted with other extensions, such as $b_j = (\tilde{b}_j +\alpha)\sigma^x_j$ with a constant $\alpha$ \cite{chamberlandBuilding2022}, for which the total Hilbert space does not form an irreducible Fock space for $b_j$.}.  The spin degrees of freedom can also be mapped to hard-core bosons, allowing a canonical boson to be expressed in  terms of a canonical $\tilde{b}$ boson and an additional hard-core boson. The above decomposition of $b$ is particularly convenient for gauging the $\bbz_2$ subgroup in the conventional manner, as in Ref.~\cite{su2024}. See Appendix~\ref{sec:bosonic_gauging} for further discussion. 
Furthermore, a fractional JW transformation $c_j=\prod_{i<j}e^{i\theta n_{\sigma,i}}\sigma_j^-$, with $n_{\sigma,i}=(1+\sigma_i^z)/2$, converts the spin degrees of freedom into hard-core anyons. This motivates the analogous composite operators $\sqrt{2}\tilde b_jc_j^\dagger+\sqrt{2\tilde n_{b,j}+1} c_j$, providing a generalization of our construction to mixtures of canonical bosons and hard-core anyons. At $\theta=\pi$, these composites reduce to the fermionic operators $A_j$.

\section{Low-energy theory} 
\label{sec:eft}
In this section, we argue that the fermionic composite $A_j$ in the fermionic dual flows to a canonical fermionic field in the low-energy theory, and that Fermi-like points determined by the total density, $\tilde{n}_j =  A_j^{\dagger} A_j =2\tilde{n}_{b, j} + n_{f, j}$, emerge in the gapless phase and satisfy a novel form of Luttinger's theorem.

Given its duality to the BH model, we expect the phases and phase transition of the fermionic dual model to be in one-to-one correspondence with those in the BH model, despite the composite nature of $A_j$. The low-energy effective theory of the BH model is described by Luttinger liquid theory \cite{gogolin2004bosonization, giamarchiQuantum2003}.   At commensurate fillings $\rho_0 = p/q$, where $p$ and $q$ are coprime integers, the (Euclidean) effective action is given by the sine-Gordon action  
\be 
S_{\rm SG} = \frac{1}{2\pi K } \int dx d\tau (\partial_{\mu} \phi)^2 + \lambda \int dx d\tau \cos  2q \phi. 
\label{eq:SG}
\ee 
Here, the sound velocity is set to unity. The real scalar field $\phi(x,\tau)$ describes the long-wavelength boson density fluctuations and, for the bosonic BH model, is compactified as $\phi\sim\phi+\pi$, with the conjugate phase field satisfying $\theta\sim\theta+2\pi$.  The first term is simply the action of the compact boson CFT. $K$ is the Luttinger parameter in the bosonization language that controls the radius of the compactification in the CFT. The second term is an interaction term that can drive the phase transition. When the cosine perturbation is irrelevant, $\phi$ fluctuates freely and the theory describes a superfluid phase. When it flows to strong coupling, $\phi$ becomes pinned at a minimum of the cosine, opening a gap and producing a commensurate insulating phase. 

A dual description of the same model is formulated in terms of the phase field $\theta$, which describes the long-wavelength phase fluctuations. We define $\phi_L = (\phi +\theta)/2$ and $\phi_R = (\phi-\theta)/2$, so  that $\phi  =\phi_L + \phi_R $ and $\theta = \phi_L - \phi_R$. These combinations are the diagonal chiral fields (at $K=1$). The charge sectors of the compact boson theory can be labeled by vertex operators 
\be V_{n, m}  = \e^{i n \phi  + i m \theta }  = e^{ i \left(n + m \right) \phi_L +i \left(n -   m\right) \phi_R  },
\ee 
where $n\in\bbz$ and $m\in\bbz$ label the vortex/winding charge and particle sectors, respectively; with our convention $b^\dagger\sim e^{-i\theta}$, the physical particle charge is $-m$. For the bosonic BH model, the local operator algebra has even $n$, consistent with $\phi\sim\phi+\pi$, while odd-$n$ operators belong to twisted sectors. In the fermionic dual, the locality condition becomes $n=m\pmod 2$: sectors with both $n$ and $m$ even contain local bosonic operators, whereas sectors with both $n$ and $m$ odd contain local fermionic operators. The remaining sectors require appropriate twists or spin structures.

Using the language of the compact boson CFT, the operators $V_{n,m}$ label the corresponding primary sectors \cite{ginsparg1988applied}\footnote{Note that one must be careful when translating between our condensed-matter convention for the compact boson CFT and the conventional string-theory notation. With our normalization, taking $\theta$ as the compact circle coordinate, the winding number is related to our quantum number $n$ by $w=n/2$. Thus, the local operator algebra of the bosonic theory, for which $n$ is even, has integer winding, while odd $n$ corresponds to twisted sectors.}. In the gapless phase, the correlation functions of these operators exhibit power-law decay, whose scaling dimensions are given by \be \Delta_{n, m} =  h_{n, m} +\bar{h}_{n, m}= \frac{1}{4}( n^2 K + m^2/K),
\label{eq:spectrum}
\ee  
where \be h_{n, m} = \frac{1}{8}  \left( n \sqrt{K } + m/\sqrt{K}\right) ^2\ee and \be  \bar{h}_{n, m} = \frac{1}{8}  \left( n \sqrt{K} - m/\sqrt{K}\right) ^2\ee are called the conformal weights of $V_{n, m}$. Note that we can also define
$s_{n, m} =  h_{n, m} -\bar{h}_{n, m}= nm/2$, which is interpreted as the spin of $V_{n, m}$. The interaction term in Eq.~(\ref{eq:SG}) has a conformal dimension $\Delta(\cos 2q \phi) =q^2K$, which becomes relevant if  $\Delta(\cos 2q \phi) <2$. Thus, at unit filling, where $q =1$, $K_c =2$ at the BKT critical point. At half filling, where $q =2$, $K_c =1/2$ cannot be reached because $K =1$ at $U \to \infty$, and the system is always gapless. 

The compact-boson CFT describing the gapless phase is equivalent, through bosonization, to the massless Thirring model. At $q=1$, the perturbation $\cos 2\phi$ in Eq.~(\ref{eq}) corresponds to a Dirac mass term, and the sine-Gordon model is therefore equivalent to the massive Thirring model (in the Euclidean signature) \cite{coleman1975}
\be
S_{\mathrm{Th}}
= \int dx\ d\tau\left[
\bar{\psi}(\slashed{\partial}+m)\psi
+\frac{g}{2}
(\bar{\psi}\gamma^\mu\psi)
(\bar{\psi}\gamma_{\mu}\psi)
\right].
\label{eq}
\ee
Here, $\psi$ is a Dirac field and $\gamma^{\mu}$, for $\mu\in\{\tau,x\}$, are two-dimensional gamma matrices satisfying $\{\gamma^{\mu},\gamma^{\nu}\}=2\delta^{\mu\nu}$. For example, we may choose $\gamma^{\tau}=\sigma^x$ and $\gamma^x=\sigma^y$. The speed $v_F$ is taken to be 1. The Luttinger parameter is related to $g$ by $K=1/(1+g/\pi)$. $K>1$ implies attractive interaction for the fermions. With the fermion normalization used here, the Dirac mass term corresponds to $\cos 2\phi$. For $q>1$, however, $\cos 2q\phi$ is not the mass term of the same Dirac fermion, but instead represents a higher-order commensurate Umklapp interaction in the massless Thirring/Luttinger theory. We therefore keep the original charge normalization and treat $\cos 2q\phi$ as an Umklapp perturbation rather than redefining the fermionic field.

The massless Thirring model provides the standard low-energy description of a spinless fermionic Luttinger liquid: the left-moving and right-moving fields near the two Fermi points form the two components of a Dirac field. Bosonization expresses this theory in terms of the conjugate fields $\phi$ and $\theta$ \cite{haldaneEffective1981, gogolin2004bosonization, giamarchiQuantum2003}, with the smooth part of the local density given by
\be \rho(x) \approx \rho_0 - \frac{1}{\pi} \partial_x \phi(x).\ee  
A conjugate phase field $\theta(y)$ satisfies    
$ [\phi(x), \partial_y\theta(y)] = i\pi \delta(x-y)$. The bosonic field is identified with 
\be 
b^{\dagger}(x) \approx e^{-i\theta(x)}   \sum_{p \in \bbz} \beta_p e^{i2p(\pi \rho_0 x -\phi(x))},
\ee 
where $\beta_p$ depend on microscopic details. Since a JW string is $e^{i \pi \sum_{i <j} n_i} \approx e^{i\pi \rho_0 x -i\phi(x)}$, the corresponding fermionic operator is identified with 
\be 
c^{\dagger}(x) \approx  e^{-i\theta(x)}   \sum_{p \in \bbz} \beta_p e^{i(2p-1) (\pi \rho_0 x -\phi(x))},
\ee
which carries a winding charge in the sine-Gordon model. 
In the earlier literature, this latter identification on a lattice is established for  spinless fermions with filling factor $0< \rho_0 <1$, since otherwise the fermions form an insulator. The corresponding Fermi momentum is  $k_F = \pi \rho_0$.  
The associated microscopic bosons are therefore hard-core bosons. In the BH model, this limit corresponds to $U  \to \infty$, where the fermions become free. 

The above correspondence between the compact-boson CFT and the massless Thirring/Luttinger theory does not depend on these microscopic filling restrictions. Turning to our fermionic dual of the BH model, the fermionic composite operators are not subject to the Pauli exclusion principle. Therefore, the microscopic identifications extend to arbitrary filling factor $\rho_0$, with the Thirring model emerging from the fermionic dual as a result of duality at low energies. Note that fermion and boson numbers in the dual theory are not separately conserved and the filling is determined by $\tilde{n}_j =  A_j^{\dagger} A_j =2\tilde{n}_{b, j} + n_{f, j}$.

Microscopic operators can be identified with CFT primary operators through their long-distance scaling behavior and symmetry quantum numbers. In the BH model,  $b^{\dagger}_j$ at low energies flows to the vertex operator $V_{0,-1}$ at leading order. In the fermionic dual of the BH model, the charge-2 operator satisfies $\tilde{b}^{\dagger}_j \approx V_{0, -2}$.  Since the JW string operator is identified with $e^{i\pi \rho_0 x}  e^{-i\phi(x)}$, then in the fermionic dual of the BH model,  $ A^{\dagger}_j \approx \beta_{0} e^{-i\pi \rho_0 j} V_{1, -1} + \beta_1 e^{i\pi \rho_0 j} V_{-1,-1}$. Here, we can perform a sublattice rotation to change the phase of $A^{\dagger}_j$ by $(-1)^j$ so that $ A^{\dagger}_j \approx \beta_{0}  V_{1, -1} + \beta_1   V_{-1,-1}$ at unit-filling $\rho_0 =1$. Since $V_{\pm 1, -1}$ are the fermionic vertex operators in the sine-Gordon/Thirring correspondence, the identification implies that the noncanonical $A^{\dagger}_j$, at leading order, flows to a linear combination of the emergent $\psi_R^{\dagger}$ and $\psi_L^{\dagger}$ fields at low energies. The spins computed using $s =nm/2$ are consistent with the particle statistics. Note that at leading order, $A_j^{\dagger 2} \sim \tilde{b}^{\dagger}_j \sim V_{-1, -1} V_{1, -1} \sim V_{0, -2}$. In a Luttinger liquid of canonical spinless fermions,  $V_{-1, -1} V_{1, -1} \sim V_{0, -2}$ is often attributed to such separated bilinears as $f_j^{\dagger}f_{j+1}^{\dagger}$  due to $f_j^{\dagger 2} =0$, but in the fermionic dual of the soft-core bosons, $V_{0, -2}$ can already be traced back to onsite operator $A_j^{\dagger 2}$. 

The above identifications can be verified numerically using DMRG by computing the scaling dimensions of these operators, as discussed below. There is, however, one important caveat concerning the scaling of  $A_j^{\dagger}$. In the fermionic dual,  as shown below, both $\langle A_i^{\dagger} (-1)^{\tilde{n}_i} A_j\rangle \sim \langle B_i^{\dagger} A_j\rangle$ and $\langle A_i^{\dagger} A_j\rangle$ scale with  $2\Delta_{1, 1}$ at fractional filling. The two correlation functions differ by a local parity insertion. It is commonly expected that such a local modification does not alter the long-distance scaling behavior, so that the two correlation functions share the same leading power law. However, a local operator insertion can cancel the leading contribution, causing the two correlations to exhibit different scaling behavior. This occurs at integer filling:  $\langle B_i^{\dagger} A_j\rangle$ continues to scale with $2\Delta_{1, 1}$, whereas $\langle A_i^{\dagger} A_j\rangle$ rapidly decays to a very small value within a few lattice spacings. Correspondingly, $\langle b_i^{\dagger} \prod_{i <k< j} (-1)^{n_{b, k}} b_j\rangle$ and $\langle b_i^{\dagger} \prod_{i \le k< j} (-1)^{n_{b, k}} b_j\rangle$ in the original BH model differ by a local boson parity $(-1)^{n_i}$. At fractional filling,  both correlation functions exhibit the same leading scaling  $2\Delta_{1, 1}$ in the Luttinger-liquid phase. At integer filling, however, it is the former that has a scaling dimension $\Delta_{1, 1}$ at leading order. To the best of our knowledge, this subtlety has not been emphasized in the existing literature.

At generic filling, the JW string induces an oscillating phase in $A_j^\dagger$, resulting in an enveloping factor $(-1)^{j-i}\cos(\pi\rho_0|j-i|+\delta)$ in the correlation functions $\langle B_i^\dagger A_j\rangle$ and $\langle A_i^\dagger A_j\rangle$ (and also $\langle f_i^\dagger f_j\rangle$), where $\delta$ is an operator-dependent phase shift. The oscillation wave vector $\pi \rho_0 \mod \pi$ can be interpreted as the Fermi momentum $k_F$ of the low-energy theory. Heuristically, the Fourier transform of $\langle A_i^{\dagger} A_j\rangle$ (at fractional filling) therefore exhibits weak singularities at the Fermi points $\pm k_F$. For canonical fermions, $k_F =0 \mod \pi$ corresponds to  a filled band.  The distinct behaviors of $\langle B_i^{\dagger} A_j\rangle$ and $\langle A_i^{\dagger} A_j\rangle$ (and also $\langle f_i^{\dagger} f_j\rangle$) at integer filling can thus be viewed as a  manifestation of this critical behavior for noncanonical fermionic operators $A_j$ and $B_j$. The relation $k_F =\pi \rho_0 \mod \pi$ can be viewed as a novel manifestation of Luttinger's theorem \cite{luttingerGroundState1960, luttingerFermi1960, yamanakaNonperturbative1997,  oshikawa2000} applied to a Bose-Fermi mixture, in which the relevant density is the sum of the bosonic and fermionic contributions, $\tilde{n}_j = A_j^{\dagger} A_j =2\tilde{n}_{b, j} + n_{f, j}$. Earlier discussions of Luttinger's theorem in different Bose-Fermi mixtures can be found in Refs.~\cite{powellDepletion2005, colemanSum2005, sachdevFermi2006}. 
  
To end this section, we briefly discuss the emergence of the two $U(1)$ symmetries and their mixed anomaly in the CFT. At $U = 0$, the free boson model has two conserved charges $Q_0 = \sum_j b_j^{\dagger} b_j$ and $Q'_1 = i \sum_j (b^{\dagger}_j b_{j+1} - b_{j+1}^{\dagger} b_j)$. Although $Q'_1$ is generally not conserved at finite $U$, these two operators flow to the particle/momentum and,  up to normalization,  winding charges, respectively, giving rise to two $U(1)$ symmetries with a mixed anomaly in the compact boson CFT. In the fermionic dual, these correspond to vector and axial charges, with the corresponding $U(1)$ symmetries exhibiting a chiral anomaly. Since the free boson model is integrable, it has infinitely many additional conserved charges. For example, in a rotated basis, $Q_1 = \sum_j (b^{\dagger}_j - b_j)(b^{\dagger}_{j+1} - b_{j+1})$ is another conserved charge. In fact, $Q_0$ and $Q_1$ generate a subalgebra of an $su(1,1)$ polynomial current algebra in the thermodynamic limit (see Appendix~\ref{sec:loopalgebbra}). In general, however, $Q_1$ is dominated by $b^{\dagger 2}$ and $b^2$ and does not flow to the vortex charge in the low-energy theory. In the hard-core limit, or equivalently in the free-fermion limit, a distinct Onsager-algebra structure exhibits an anomaly that flows to the chiral anomaly in the low-energy theory \cite{chatterjee2025quantized, pace2024lattice, lew-smithInfiniteOrder2026}.

\begin{figure}[tb]
    \centering
    \includegraphics[width=1.0\linewidth]{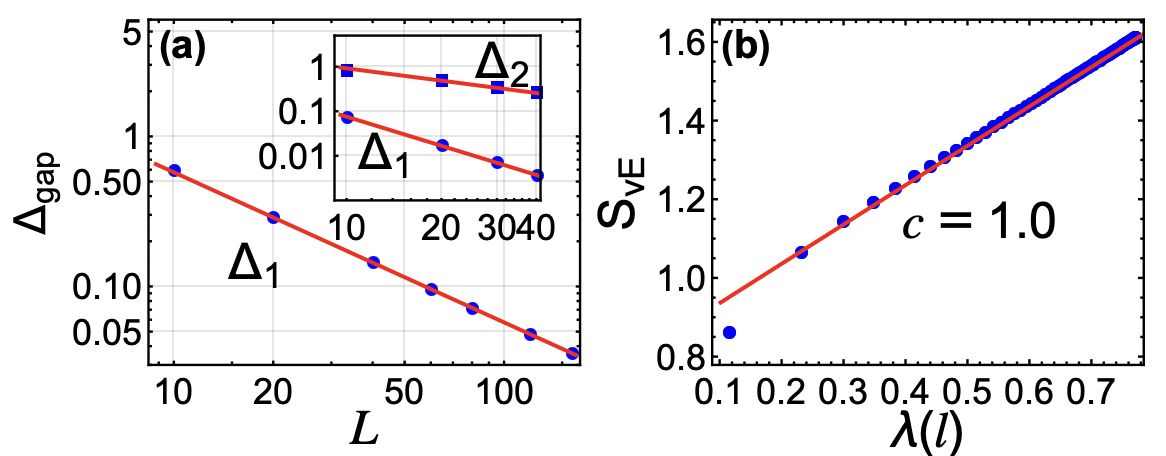}
    \caption{(a) Finite-size scaling of $\Delta_1 = E_1 - E_{gs}$ in the gapless phase of the fermionic dual at unit-filling for $U/t=2.0$ under OBCs. The slope of the fit is $-1$. Inset: finite-size scaling of $\Delta_1 = E_1 - E_{gs}$ and $\Delta_2 = E_2 - E_{gs}$ under PBCs with fitted slopes $-2.15$ and $-0.89$, respectively. $\Delta_1$ decreases faster than $1/L$, signaling a twofold ground-state degeneracy in the thermodynamic limit.  (b) Subsystem entanglement entropy as a function of $\lambda(l) = \frac{1}{6} \ln(\frac{2L}{\pi} \sin \frac{\pi l}{L})$ for $U/t=2.0$ under OBCs, where $l$ is the subsystem size and $L =160$ is the total system size. The slope yields a central charge $c \approx 1.0$, consistent with the underlying CFT.}
    \label{fig2} 
\end{figure}

\section{Numerics}
\label{sec:numerics}
In this section, we verify the duality between the BH model and its fermionic dual in Eq.~(\ref{eq:dualBH}) using DMRG calculations, implemented with the ITensor package \cite{itensor}. We compute the spectra of both models under different boundary conditions and determine the central charge of the gapless phase of the fermionic dual. More importantly, we extract the scaling dimensions of the correlation functions of several microscopic operators in the gapless phase and compare them with the predicted scaling dimensions of the corresponding vertex operators identified above in Luttinger liquid theory, or equivalently, the compact boson CFT. We also extract the oscillating wave vector of the fermionic composite correlation function to justify a generalized Luttinger's theorem. 
In our DMRG calculations, the local bosonic Hilbert space is truncated at occupation number $n_b \le 6$ for the BH model and  $\tilde{n}_b\le 6$ for the fermionic dual. We have verified that further increasing these cutoffs does not appreciably affect the reported results.

\subsection{Spectra  and central charge}

We first examine the spectra under different boundary conditions to confirm the expected correspondence between the original BH model and our fermionic dual. Under OBCs, the spectrum of the fermionic dual agrees within numerical accuracy with that of the original BH model, with the critical point remaining at $U/t \approx 3.3$. In this gapless phase of the fermionic dual, as an example,  the first finite-size excitation gap, which agrees within numerical accuracy with that of the BH model, closes as $1/L$, as shown in Fig.~\ref{fig2}(a). In the following, we focus on the gapless phase.

Under periodic  boundary conditions (PBCs), the ground state of the BH model is unique. From the CFT perspective, this unique state corresponds to $n = m =0$ in Eq.~(\ref{eq:spectrum}). Under antiperiodic boundary conditions (APBCs), the phase field $\theta$ must wind by an odd multiple of $\pi$ across the length of the system. In our convention, the corresponding winding quantum number $n$ therefore takes odd-integer values, with the two lowest-energy sectors $n=\pm1$ giving a doubly degenerate ground state.  See Refs.~\cite{cheng2023lieb, su2025z2} for further discussion in this context. Numerically, the two-fold degeneracy is exact at fractional filling for finite systems. At integer filling, the first excitation gap $\Delta_1$ closes as $L^{-\alpha}$ with $\alpha >1$, signaling a ground-state degeneracy in the thermodynamic limit. 

These behaviors are also observed in the fermionic dual. Imposing periodic (Ramond) boundary conditions (PBCs) at unit filling, we find that for even $L$ (and thus even parity) the first gap $\Delta_1$ closes as $L^{-\alpha}$ with $\alpha \sim 2$ while the second gap $\Delta_2$ scales as $L^{-\alpha}$ with $\alpha \sim 1$, as shown in the inset of Fig.~\ref{fig2}(a). This indicates a two-fold degeneracy in the thermodynamic limit. In contrast, for even fermion parity, imposing antiperiodic (Neveu-Schwarz) boundary conditions (APBCs) does not lead to such a degeneracy.   This behavior is consistent with that of the BH model. More generally, the bosonic and fermionic boundary signs are related by $\eta_b=-(-1)^N\eta_f$. Thus, for even particle number the PBC and APBC sectors are interchanged, whereas for odd particle number they coincide. Indeed, for odd $L$, corresponding to odd parity, the dependence of the ground-state degeneracy of the fermionic dual on the boundary conditions coincides with that of the BH model. These results provide a confirmation that the dual theory is genuinely fermionic.

As a further confirmation of the duality, we compute the central charge $c$ in the gapless phase of the fermionic dual. As shown in Fig.~\ref{fig2}(b), the subsystem entanglement entropy  $S_{vE} = -\tr(\rho_l \ln \rho_l)$, where $\rho_l$ is the reduced density matrix of a subsystem, scales linearly with the factor $\lambda(l)\equiv \frac{1}{6}\log \left(\frac{2L}{\pi}\sin\left(\frac{\pi l}{L}\right)\right)$ \cite{calabrese2009entanglement}.  Here, $L$ is the total system size of the open chain, and $l$ is the size of the subsystem on one side of the bipartition. The resulting slope, $c \approx 1.0$, agrees with that of the  superfluid phase of the BH model, and identifies the low-energy theory as the compact-boson CFT with central charge $c =1$.
 
\subsection{Correlation functions}

Now we turn to the scaling of correlation functions. Since the fermionic correlations including $\langle A_i^{\dagger} A_j \rangle $ and $\langle f_i^{\dagger} f_j \rangle$ exhibit distinct behavior at fractional and integer fillings, we consider two representative cases: half filling and unit filling.

\begin{figure}[tb]
    \centering
    \includegraphics[width=1.0\linewidth]{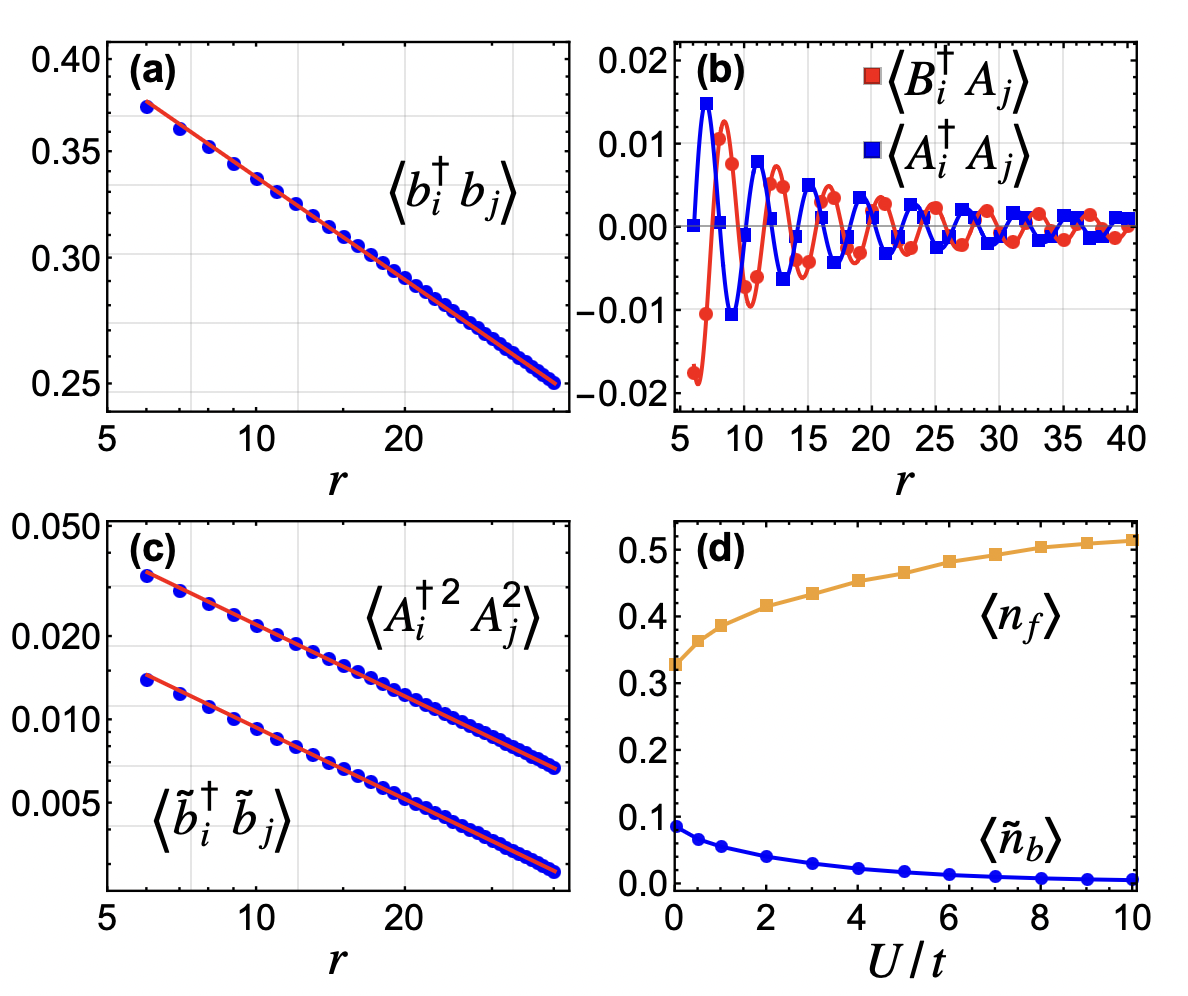}
    \caption{(a) Scaling of $\langle b_i^{\dagger} b_j \rangle $ in the superfluid phase of the BH model at half filling, $\langle n_b\rangle =0.5$, as a function of $r = |j-i|$, for $U/t=2.0$ and $L =160$ under OBCs. The starting site is $i =L/4$.  The fitted slope yields $2\Delta_{0, 1} \approx 0.215$, corresponding to $K \approx 2.3$ in Eq.~(\ref{eq:spectrum}).  (b-c) Gapless phase of the fermionic dual at half filling, $\langle 2 \tilde{n}_b + n_f \rangle =0.5$. (b) $\langle B_i^{\dagger} A_j \rangle $ and $\langle A_i^{\dagger} A_j \rangle $ for $U/t=2.0$ and $L =160$ under OBCs, both fitted to $a \sin(\pi \rho_0 r +b) r^{-2\Delta_{1,1}}$, with $2\Delta_{1, 1} \approx 1.38$ and $\rho_0 \approx 0.49$. The correlation $\langle f_i^{\dagger} f_j \rangle$ (not shown) exhibits similar behavior to $\langle A_i^{\dagger} A_j \rangle $. (c) $\langle A_i^{\dagger 2} A_j^2 \rangle $ and $\langle \tilde{b}_i^{\dagger} \tilde{b}_j \rangle $ shown on a log-log scale, with the fits determined by $2\Delta_{0, 2} \approx 0.86$. (d) Average onsite densities $\langle \tilde{n}_{b} \rangle $ and $\langle n_f \rangle$ as a function of $U/t$ under PBCs for $L =20$. As $U/t$ increases, $A_j$ approaches the hard-core, or equivalently free-fermion, limit.}
    \label{fig3}
\end{figure}

\begin{figure}[tb]
    \centering
    \includegraphics[width=1.0\linewidth]{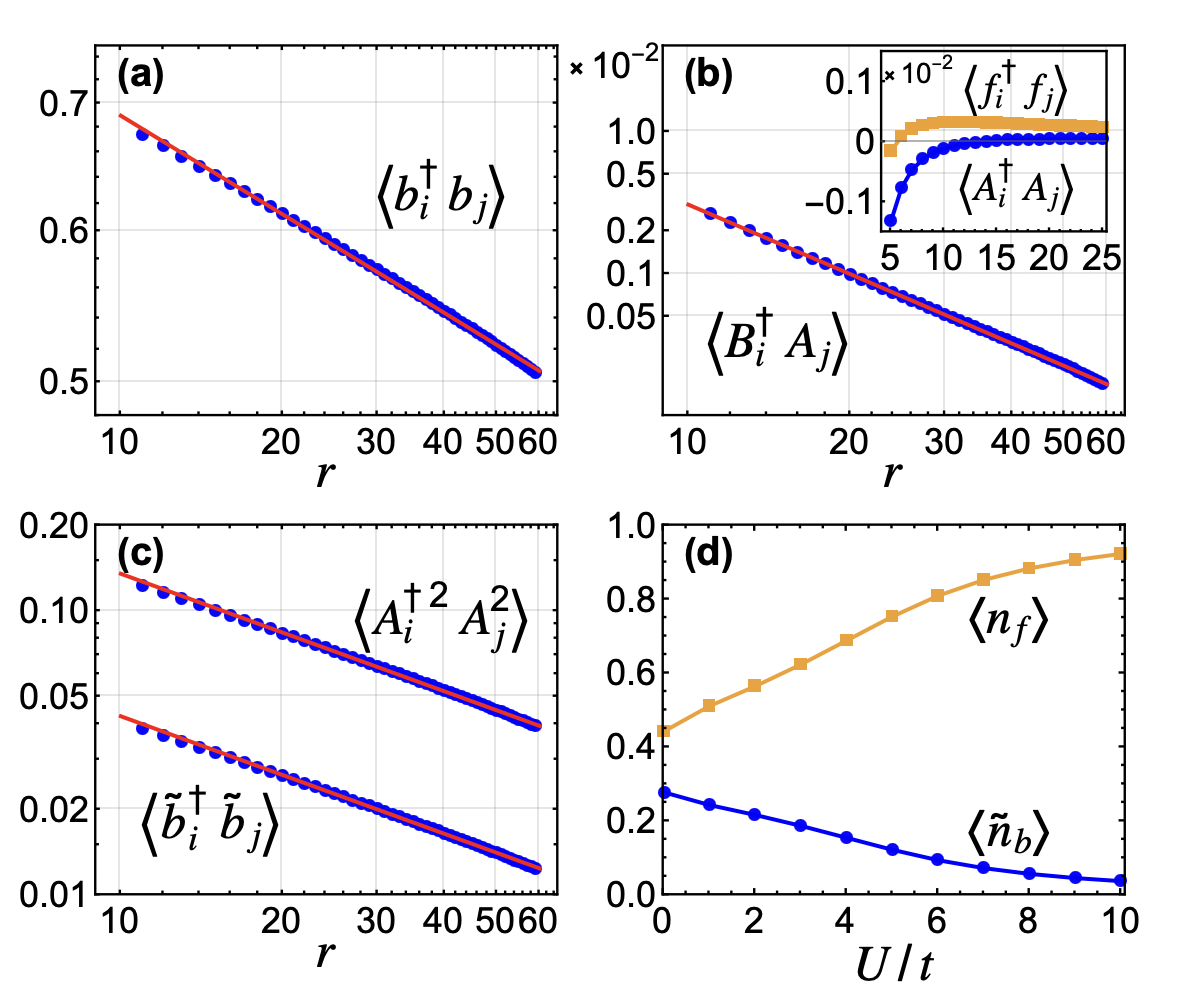}
    \caption{
    (a) Scaling of $\langle b_i^{\dagger} b_j \rangle $ in the superfluid phase of the BH model at unit filling, $\langle n_b\rangle =1$, as a function of $r = |j-i|$, for $U/t=2.0$ and $L =160$ under OBCs. The starting site is $i =L/4$.  The fitted slope yields $2\Delta_{0, 1} \approx 0.172$, corresponding to $K \approx 2.9$ in Eq.~(\ref{eq:spectrum}). (b-c) Gapless phase of the fermionic dual at unit filling, $\langle 2 \tilde{n}_b + n_f \rangle =1$.  (b) $\langle B_i^{\dagger} A_j \rangle $ for $U/t=2.0$ and $L =160$ under OBCs, shown on a log-log scale, with the fit determined by $2\Delta_{1, 1} \approx 1.63$. Inset: scaling of $\langle f_i^{\dagger} f_j \rangle $ and $\langle A_i^{\dagger} A_j \rangle $, both of which decay rapidly to near zero within a few lattice spacings. (c) $\langle A_i^{\dagger 2} A_j^2 \rangle $ and $\langle \tilde{b}_i^{\dagger} \tilde{b}_j \rangle $ shown on a log-log scale, with the fits determined by $2\Delta_{0, 2} \approx 0.69$. (d) Average onsite densities $\langle \tilde{n}_{b} \rangle $ and $\langle n_f \rangle$ as a function of $U/t$ under PBCs for $L =20$. The critical point is $U/t \approx 3.3$. As $U/t$ increases, $A_j$ approaches $f_j$.}
    \label{fig4}
\end{figure}

\subsubsection{Half filling}
The fermionic dual in this case is always in the gapless phase for finite $U$.  As a benchmark, we extract the scaling dimension $\Delta_{0,1}$ from $\langle b^{\dagger}_i b_j\rangle \sim r^{-2\Delta_{0, 1}}$ with $r =|j-i|$ in the original BH model, as shown in Fig.~\ref{fig3}(a), for $U/t=2.0$ and $L =160$. To reach larger system sizes, we use OBCs. The correlation function is computed within a symmetric interval deep in the bulk, centered around the middle of the chain, to minimize boundary effects. This yields a Luttinger parameter $K \approx 2.3$, which we then use in Eq.~(\ref{eq:spectrum}) to determine the corresponding scaling dimensions. We compare these predictions with the scaling dimensions extracted by fitting the correlation functions in the fermionic dual. Strictly speaking, $K$ flows with the length scale under renormalization; here, we work at a fixed finite size and use the corresponding effective value of $K$ to test the CFT predictions of Eq.~(\ref{eq:spectrum}).   

We present the oscillating correlation functions $\langle B_i^{\dagger} A_j \rangle $ and $\langle A_i^{\dagger} A_j \rangle $ in Fig.~\ref{fig3}(b). As discussed in the previous section, the two correlations differ by a local parity operator. At fractional filling, they exhibit similar scaling behavior, of the form:  $a \sin(\pi \rho_0 r +b) r^{-2\Delta_{1,1}}$. The extracted power-law exponent is close to the predicted value $2\Delta_{1, 1} \approx 1.38$, while the oscillation frequency $\rho_0 \approx 0.49$ is close to the half-filling value $\rho_0 =0.5$. The correlation $\langle f_i^{\dagger} f_j \rangle$, not shown in Fig.~\ref{fig3}(b), exhibits similar behavior to $\langle A_i^{\dagger} A_j \rangle $. We also present $\langle A_i^{\dagger 2} A_j^2 \rangle $ and $\langle \tilde{b}_i^{\dagger} \tilde{b}_j \rangle $ in Fig.~\ref{fig3}(c). The former can be viewed as a generalized Cooper-pair correlation. The extracted scaling dimensions again agree with $2\Delta_{0, 2} \approx 0.86$.  In addition, throughout the Luttinger-liquid phase, we confirm that the operator $Q' = -i\sum_j (B_j^{\dagger} A_{j+1} -A_{j+1}^{\dagger} B_j)$ flows to the winding charge whose local density has a scaling dimension $\Delta =1$.

In the large-$U$ limit, both $\langle B_i^{\dagger} A_j \rangle $ and $\langle A_i^{\dagger} A_j \rangle $ reduce to $\langle f_i^{\dagger} f_j \rangle$, up to a sign. At finite $U$, interactions suppress high-occupancy states in the low-energy sector and modify the scaling behavior, but in such a way that the noncanonical fermionic composites $A_j$ and $B_j$, as well as the canonical fermion $f_j$, flow to the emergent dressed fermionic fields $\psi$ and $\bar{\psi}$ (up to normalizations) in the low-energy theory. This correspondence is particularly transparent for  $\rho_0 <1$, where the composite of the soft-core boson $\tilde{b}_j$ and fermion $f_j$ can be  viewed as a smooth deformation of the hard-core boson, or equivalently the free fermion $f_j$. For $\rho_0 >1$, the magnitudes  of $\langle B_i^{\dagger} A_j \rangle $ and $\langle A_i^{\dagger} A_j \rangle $ become very small while
$\langle f_i^{\dagger} f_j \rangle$ is suppressed even further. At large filling, $f_j$ fermions dressed with bosons have larger correlation functions, as in $\langle \sqrt{2}\tilde{b}^{\dagger}_{i} f_{i} \sqrt{2}\tilde{b}_{j} f^{\dagger}_{j} \rangle$ and $\langle \sqrt{2 \tilde{n}_{b, i} +1} f_{i}^{\dagger} \sqrt{2 \tilde{n}_{b, j} +1} f_{j}\rangle$. Nevertheless, our numerical results indicate that oscillatory behavior persists in $\langle f_i^{\dagger} f_j \rangle$ even in this regime. 

The composite nature of $A_j$ (and $B_j$) can be elucidated by tracking the evolution of the occupancies $\langle \tilde{n}_b\rangle$ and $\langle n_f\rangle$ as a function of $U/t$. At half filling, the constraint $2\langle \tilde{n}_b\rangle  + \langle n_f \rangle = \langle \tilde{n} \rangle =0.5$ must be satisfied. As the interaction strength increases, $\langle n_f\rangle$  increases toward 0.5 while $\langle \tilde{n}_b\rangle$  decreases smoothly toward 0, as shown in Fig.~\ref{fig3}(d). Over a substantial range of $U/t >0$, we find $\langle\tilde{n}_b\rangle \ll \langle n_f \rangle$, indicating that $A_j$ has a substantial overlap with the canonical fermion $f_j$. As $U$ increases, $A_j$, at low energies, continuously approaches the canonical fermionic operator  $f_j$ and excitations created by $\tilde{b}^{\dagger}$ are suppressed.  At $U/t =2.0$, $\langle \tilde{n}_{b, j} \rangle \ll 1$, and a simple mean-field estimate gives $A_j^{\dagger 2} = (\sqrt{2}\tilde{b}_{j}^{\dagger} f_{j} + \sqrt{2 \tilde{n}_{b, j} +1} f_{j}^{\dagger})^2 \sim 1.9\cdot \tilde{b}^{\dagger}_j$. This estimate is roughly reflected in the relative magnitudes of the correlation functions  $\langle \tilde{b}^{\dagger}_i \tilde{b}_j\rangle$ and $\langle A^{\dagger 2}_i A_j^2 \rangle$ in Fig.~\ref{fig3}(c).

\subsubsection{Unit filling} 
We now turn to the unit-filling case. As in the half-filling case, we first extract the scaling dimension $\Delta_{0, 1} \approx 0.086$ from $\langle b^{\dagger}_i b_j\rangle $ [shown in Fig.~\ref{fig4}(a)] in the original BH model for $U/t =2.0$ and $L =160$ under OBCs. This gives $K \approx 2.9$ through the relation $\Delta_{n, m}$ in Eq.~(\ref{eq:spectrum}).  Using the same value of $K$ in the fermionic dual, we find that the correlations $\langle B^{\dagger}_i A_j\rangle \sim r^{-2\Delta_{1, 1} }$, $\langle A^{\dagger 2}_i A_j^2 \rangle \sim r^{-2 \Delta_{0,2}}$, and $\langle \tilde{b}^{\dagger}_i \tilde{b}_j\rangle \sim r^{-2 \Delta_{0,2} }$, all exhibit scaling dimensions consistent  with the  CFT predictions, as shown in Fig.~\ref{fig4}(b,c). The magnitude of $\langle B^{\dagger}_i A_j\rangle$ is substantially smaller than those of the other two correlations because of its larger scaling dimension implies a faster asymptotic decay than those correlations, as is evident from Eq.~(\ref{eq:spectrum}) upon substituting the extracted value of $K$. In contrast, $\langle f_i^{\dagger} f_j\rangle$ rapidly decays to near zero within a few lattice spacings, as shown in the inset of Fig.~\ref{fig4}(b). Numerics show that the dominant contribution to the algebraic scaling of $\langle B^{\dagger}_i A_j\rangle$ arises from terms of the form $\langle  \tilde{b}^{\dagger}_{i} f_{i} f_{j} \rangle  \approx - \langle  \tilde{b}_{j} f_{i}^{\dagger} f_{j}^{\dagger} \rangle $, in which an $f$ fermion is annihilated at one site while another is annihilated at another site together with the creation of a $\tilde{b}$ boson. Here, we omit the factors $ \sqrt{2 \tilde{n}_{b, j} +1}$ for simplicity. The other two contributions in the expansion, of the forms $\langle \tilde{b}_i^{\dagger} \tilde{b}_j f_i f_j^{\dagger}\rangle$ and $\langle f_i^{\dagger} f_j \rangle$, as given in Eq.~(\ref{eq:dualBH}), decay algebraically at fractional filling but are more strongly suppressed at integer filling and therefore decay faster than the leading power law.
Because of the relative sign between the two terms,  $\langle  \tilde{b}^{\dagger}_{i} f_{i} f_{j} \rangle \approx- \langle  \tilde{b}_{j} f_{i}^{\dagger} f_{j}^{\dagger} \rangle $, their contributions to  $\langle A^{\dagger}_i A_j\rangle$ cancel.  Consequently, as shown in the inset, $\langle A^{\dagger}_i A_j\rangle$  also rapidly crosses zero within a few lattice spacings at unit filling. This change in behavior can also be viewed as the underlying mechanism responsible for the different scaling of $\langle b_i^{\dagger} \prod_{i <k< j} (-1)^{n_{b, k}} b_j\rangle$ and $\langle b_i^{\dagger} \prod_{i \le k< j} (-1)^{n_{b, k}} b_j\rangle$ in the original BH model, as mentioned in the previous section. 

Consistent with a generalized Luttinger's theorem, the integer-filling case should be viewed as a limiting case of the fractional-filling regime. For canonical spinless fermions with conserved particle number, the system is insulating at unit filling. More generally, when lower bands, if present, are completely filled, the system is insulating at any integer filling. In our composite system, canonical fermions $f$  reside on top of the canonical bosonic levels of $b$, labeled by $n_b$, or equivalently of $\tilde{b}$, labeled by $\tilde{n}_b = n_b/2$, as illustrated in Fig.~\ref{fig1}. However, because the canonical fermion number is not separately conserved, heuristically, the effective fermion density is given by the density of the fermion composites, $\tilde{n}_j =  A_j^{\dagger} A_j =2\tilde{n}_{b, j} + n_{f, j}$. At integer filling, the system can remain gapless, but correlation functions such as $\langle \tilde{b}_i^{\dagger} \tilde{b}_j f_i f_j^{\dagger}\rangle$ and $\langle f_i^{\dagger} f_j \rangle$,  which  separately preserve the numbers of $\tilde{b}$ bosons and $f$ fermions, can detect the change in the filling.  Our heuristic picture is motivated by numerical observations and calls for a more comprehensive theoretical understanding in future work.

As $U$ is increased, the system goes through a BKT transition into a Mott-insulating phase. At the critical point, $U/t\approx 3.3$, the Luttinger parameter reaches $K \approx 2.0$, in agreement with the corresponding value in the BH model. As in the half-filling case, we elucidate the composite nature of $A_j$  by examining the evolution of the occupancies $\langle \tilde{n}_b\rangle$ and $\langle n_f\rangle$ as a function of $U/t$. The unit-filling constraint requires $2\langle \tilde{n}_b\rangle  + \langle n_f \rangle = \langle \tilde{n} \rangle =1$. As the interaction strength increases, $\langle n_f\rangle$  increases smoothly toward 1 while $\langle \tilde{n}_b\rangle$ decreases smoothly  toward 0, as shown in Fig.~\ref{fig4}(d), with no discernible discontinuity at the critical point. In the gapless phase for $U/t < 3.3$, $\langle \tilde{n}_{b, j} \rangle \sim 0.25$, suggesting $A_j^{\dagger 2} \sim 2.2\cdot \tilde{b}^{\dagger}_j$. The relative magnitudes of the correlation functions $\langle A^{\dagger 2}_i A_j^2 \rangle$ and  $\langle \tilde{b}^{\dagger}_i \tilde{b}_j\rangle$ are of the same order as suggested by this estimate, as shown in Fig.~\ref{fig4}(c). 

At other integer fillings, the above analysis remains qualitatively unchanged, except that $\langle B^{\dagger}_i A_j\rangle$ may acquire an additional enveloping factor $(-1)^r\cos(\pi \rho_0 r)$. For example, at $\rho_0 =2$, the correlation function exhibits oscillations with a period of two lattice spacings. Note that, at even integer fillings, this staggering can be removed by undoing the sublattice rotation, $A_j\to(-1)^jA_j$ and $B_j\to(-1)^jB_j$.

\subsection{Edge-to-edge correlations}
Note that when we compute correlation functions under OBCs, we choose a symmetric interval sitting around the center of the chain to minimize effects from boundaries. When $U/t$ is small, there seem to be nonvanishing  revivals in the edge-to-edge correlation function $\langle A^{\dagger}_1 B_L\rangle$ (or $\langle B^{\dagger}_1 A_L\rangle$), even when $L$ is large (see  Appendix~\ref{sec:edgetoedge} for plots and additional discussion). There are some recent observations of   edge-like features associated with the descendant of the mean-field Majorana edge mode in the extended spinless Fermi-Hubbard model with nearest-neighbor attraction \cite{thomas-markarianMajorana2025, debortoliMajorana2026} (see also Ref.~\cite{yinMajorana2019}).  However, these observations may be a finite-size effect, since the finite-size gap closes as $1/L$ under OBCs. Indeed, $\langle A^{\dagger}_1 B_L\rangle$ and $\langle b^{\dagger}_1 b_L\rangle$, which are related by a JW transformation, are identical up to a sign determined by the  total particle parity when the total particle number is fixed. Thus, an algebraic decay  in $\langle b^{\dagger}_1b_r\rangle$ rules out the possibility of nonvanishing  $\langle A^{\dagger}_1 B_L\rangle$ in the thermodynamic limit. Nevertheless, when $U/t \to 0$,  $K\to \infty$ and $\Delta(b) \to 0$, allowing the revival to persist even for large system sizes.   

\section{2D generalization}
\label{sec:2dgen}
Following the 1D case, one may consider 2D Hamiltonians in terms of $A_j = \sqrt{2}\tilde{b}_{j} f^{\dagger}_{j} + \sqrt{2 \tilde{n}_{b, j} +1} f_{j}$.
However, for the purpose of this work, we apply fermionic gauging to the BH model following Ref.~\cite{su$mathbbZ_2$2025a}, and derive the fermionic dual of the 2D BH model. In 1D, fermionic gauging is equivalent to a slightly generalized JW transformation. Its advantages become more apparent in higher dimensions, where it provides a natural generalization of the JW transformation. The construction applies to general lattices; here, we focus on the square and triangular lattices.

\subsection{Square lattice} 
\begin{figure}[tb]
    \centering
    \includegraphics[width=1.0\linewidth]{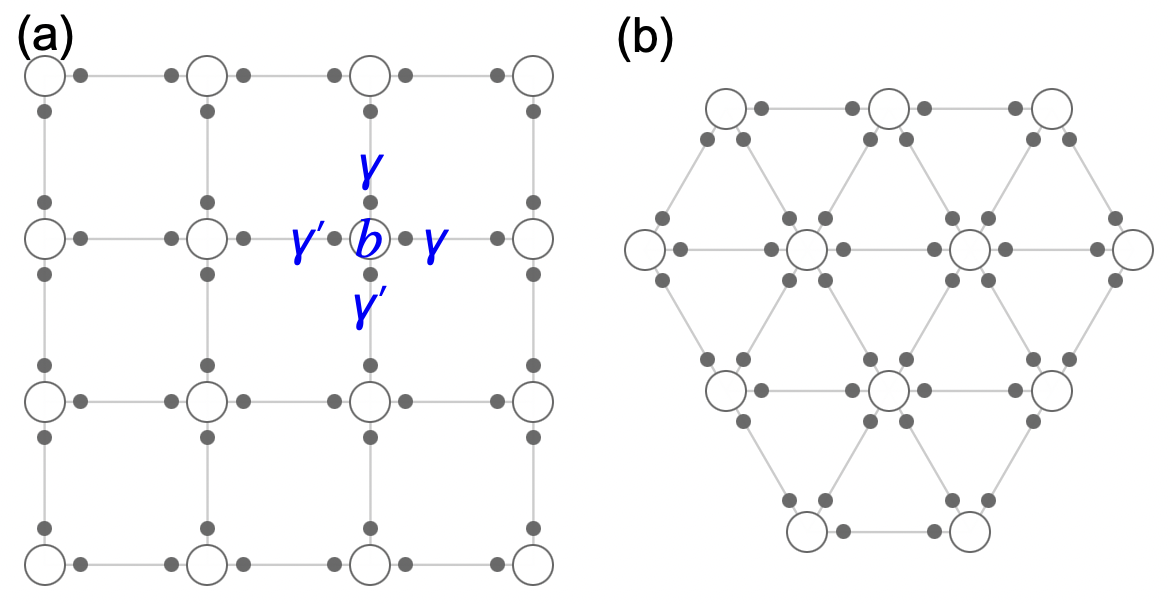}
    \caption{Fermionic gauging of the 2D Bose-Hubbard model on the square lattice (a) and triangular lattice (b). One $b$ boson is placed on each vertex and a pair of Majorana fermions, $\gamma$ and $\gamma'$, is introduced on each edge such that each boson is surrounded by four (six) Majorana fermions. A Gauss law is imposed at each vertex, while a flatness condition is imposed on the eight (six) Majorana fermions surrounding each square (triangular) face. }
    \label{fig6}
\end{figure}

The Hamiltonian of the 2D BH model on the square lattice [Fig.~\ref{fig6}(a)] is 
\be 
H = -t \sum_{\langle i, j\rangle } ( b^{\dagger}_i b_{j} + b_i b_{j}^{\dagger})   + \frac{U}{2} \sum_j n_j  (n_j -1),  
\ee
where $\langle i, j\rangle$ represent nearest-neighbor sites. We write down the minimally coupled Hamiltonian    
\be 
\begin{split}
H' = & -t \sum_{\langle i, j\rangle } [ b^{\dagger}_i (-1)^{n_{f, e_{ij}}} b_{j} + b_i (-1)^{n_{f, e_{ij}}} b_{j}^{\dagger}]  \\ & + \frac{U}{2} \sum_j n_j  (n_j -1).   
\end{split}
\ee 
Here, we generalize the 1D construction by placing a pair of Majorana fermions on each edge $e_{ij}$ and assigning the two Majorana fermions to the neighboring sites $i$ and $j$ connected by that edge. Thus, on the square lattice, each site $j$ is associated with four Majorana fermions, which we denote by $\gamma_j^e$. To distinguish the two Majorana fermions on each edge, we also use  $\gamma$ and $\gamma'$, respectively, to label the Majorana fermion to the left/right or bottom/top on the same edge such that $(-1)^{n_{f, e}} = i \gamma_e \gamma'_e$. Here, $\gamma'$ should be identified with some $\gamma_j^e$. 

The Gauss law at each vertex $j$ is  
\be  
(-1)^{n_{b, j}} i^2 \prod_{\partial e \supset j} \gamma_j^e =1.
\ee Here, a fixed ordering of $\gamma_j^e$ is implicitly chosen in the product, which can also be written as the fermion parity for two complex fermions. To preserve the duality, we need to impose the flatness condition on each face $p$:
\be 
\prod_{ e \subset \partial p} (-1)^{n_{f, e}} = 1,
\ee 
which involves eight surrounding Majorana fermions. 
The disentangling unitary is given by 
\be
\tilde{U} = \prod_j(\Pp_{j}^+ +\Pp_{j}^- K_j),
\ee 
where $\Pp_{j}^{\pm} = [1\pm i^2 \prod_{\partial e \supset  j} \gamma_j^e]/2$ and $K_j$ is the onsite parity-switching operator as in the 1D case. Under the unitary transformation, the Gauss law is mapped to $(-1)^{n_{b, j}} =1$, and the local Hilbert space sectors of odd boson occupancy are projected out. Moreover,
\be 
n_{b, j} \to  n_{b, j} + \Pp_{j}^-, \quad b^{\dagger}_i (-1)^{n_{f, e_{ij}}} b_{j} \to 
   \tilde{A}_i^{\dagger} (-1)^{n_{f, e_{ij}}} \tilde{A}_j,  
\ee  
where $\tilde{A}_j = \sqrt{2} \tilde{b}_j\Pp_j^+ + \sqrt{2\tilde{n}_{b, j} +1} \Pp_j^-$ with $ \tilde{b}_j = [2(n_{b,j}+1)]^{-1/2} b_j^{2}$, the same as in 1D.   $\tilde{A}_j$ can be multiplied by the Majorana fermion at site $j$ in $ (-1)^{n_{f, e_{ij}}}$ and becomes fermionic.  The flatness condition remains invariant. 
Thus, the exact dual Hamiltonian is 
\be  
H''  =  -t\sum_{\langle i, j\rangle } \left[\tilde{A}_i^{\dagger} (-1)^{n_{f, e_{ij}}} \tilde{A}_j + h.c.\right]    
 + \frac{U}{2} \sum_j \tilde{n}_j(\tilde{n}_j-1)
 \label{eq:dualBH2D}
\ee 
with a fermionic analog of the new effective Gauss law $\prod_{ e \subset \partial p} (-1)^{n_{f, e}} = 1$. Here,  $\tilde{n}_j = \tilde{A}^{\dagger}_j \tilde{A}_j  =  2\tilde{n}_{b, j} + \Pp_{ j}^- $. Note that, due to the effective Gauss law, fermionic point-like excitations are prohibited in order to preserve the exact duality, in sharp contrast to the 1D case.

An alternative way to derive the above dual model is to apply fermionic gauging directly to the Pauli spin operators in the boson parity decomposition $b_j \equiv   \sqrt{2}\tilde{b}_{j} \sigma^+_{j}  +  \sqrt{2 \tilde{n}_{b, j} +1} \sigma^-_{j} $ discussed in Appendix~\ref{sec:decomposition}. In the hard-core limit, the gauging procedure essentially reduces to that discussed in  Ref.~\cite{su$mathbbZ_2$2025a}, and the dual Hamiltonian is given by 
\be   
 H''_{\text{h.c.}}  =  -t\sum_{\langle i, j\rangle } \left[\Pp_i^- (-1)^{n_{f, e_{ij}}}   \Pp_j^- + h.c.\right]  
\ee 
with $\prod_{ e \subset \partial p} (-1)^{n_{f, e}} =1$. The $U(1)$ symmetry is generated by $\sum_j \Pp_{ j}^- $. A tight-binding model of hard-core bosons with hopping only in 2D is argued to be equivalent to free complex fermions coupled to a Chern-Simons theory using the flux-attachment approach \cite{fradkin1989jordan, eliezerAnyonization, eliezerIntersection1992}. Here, the dual is expressed in terms of Majorana fermions with a generalized Gauss law. Conversely, a tight-binding model of free complex fermions is dual to a gauged theory of hard-core bosons once we identify them with spin-1/2 degrees of freedom \cite{suBosonization2026, chen2018exact}. 

As in 1D, the spectral correspondence between the dual systems requires an appropriate matching of boundary conditions and global sectors. Their ground-state degeneracies can differ when all gauge sectors are retained. We therefore expect corresponding phases and phase transitions in the 2D BH model and its fermionic dual. In particular, the superfluid phase, the Mott insulating phase, and the commensurate transition governed by the Wilson–Fisher fixed point have dual realizations in a system of bosons and fermions, with observables related by the duality map. In the gapless phase, $\tilde b$ bosons condense, and we may make the mean-field approximation $\tilde b\approx\alpha$. Then $\tilde A_j\approx\sqrt{2}\alpha\Pp_j^++\sqrt{2|\alpha|^2+1}\Pp_j^-$ and $\tilde n_j\approx2|\alpha|^2+\Pp_j^-$. Within this approximation, the Hamiltonian in Eq.~(\ref{eq:dualBH2D}) is expressed entirely in terms of Majorana fermions. Although $\tilde b$ number is not separately conserved, the phase of $\alpha$ can be chosen through a global $U(1)$ transformation acting jointly on the bosonic and fermionic degrees of freedom. A gauge-invariant charge-2 pairing order parameter is $\langle(\gamma_j^e\tilde A_j)^2\rangle=\alpha'$, which is independent of the choice of incident edge $e$. At the critical point, an emergent particle-vortex duality is expected to arise \cite{seibergduality2016, karch2016}. Its implications for the fermionic dual remain to be understood in greater detail.

We emphasize that, in this section, we have used the term ``fermionic" in a loose sense, referring only to the presence of fermionic operators in the Hamiltonian, and not implying that the low-energy theory contains point-like fermionic excitations or depends on a choice of spin structure. In this sense, our usage is analogous to that of fermionization in dimensions higher than $(1{+}1)$D, such as the flux-attachment approach to fractional quantum Hall systems \cite{fradkin2013field, seibergduality2016}.  To make the physics of Eq.~(\ref{eq:dualBH2D}) more interesting,  we can relax the constraint imposed by the Gauss law, $\prod_{ e \subset \partial p} (-1)^{n_{f, e}} = 1$, and instead enforce it energetically. Fermionic point-like excitations and other anyonic excitations are then allowed, analogous to those in the toric code. We briefly discuss this aspect below on the triangular lattice.

\subsection{Triangular lattice}
The new Gauss law on the square lattice involves eight Majorana fermions. To reduce the weight of fermion numbers, we can consider the triangular lattice as in Fig.~\ref{fig6}(b) and repeat the same procedure. The corresponding Majorana fermions form a honeycomb lattice. Then both the projection operator $\Pp_{j}^{\pm} = [1\pm i^3 \prod_{\partial e \supset  j} \gamma_j^e]/2$ and the Gauss law $\prod_{ e \subset \partial p} (-1)^{n_{f, e}} = 1$ involve six Majorana fermions. 

If the $b$ bosons are replaced by Ising spins, as in Ref.~\cite{su$mathbbZ_2$2025a}, the corresponding fermionic dual, after relaxing the Gauss law, can be a Majorana fermion surface code \cite{bravyiMajorana2010}, generalizing qubit stabilizer codes, such as the toric code \cite{kitaev2006anyons}. These codes have been extensively studied in recent years, especially on the honeycomb lattice \cite{vijay2015majorana} (see, e.g., Ref.~\cite{terhal2012from} for a version on the square lattice). Majorana stabilizer codes can exhibit some advantages over their qubit counterparts. In Appendix \ref{sec:Maj}, we provide more details about these codes from a gauging perspective.
Fermionic gauging can also be applied to the nontrivial Ising paramagnet in 2D to obtain a Majorana commuting-projector model whose excitations are topologically distinct from those of the Majorana fermion surface code, generalizing the Levin-Gu diagnostic for symmetry-protected topological (SPT) phases \cite{levin2012braiding}. 

A fermionic dual obtained by fermionically gauging the $\bbz_2$ subgroup of the $U(1)$ symmetry of a system of $b$ bosons can be viewed as a generalization of the Majorana fermion surface code,  which is obtained by gauging the $\bbz_2$ symmetry of an Ising model. The Majorana fermion surface code is formulated entirely in terms of Majorana fermions, whereas the generalized fermionic duals involve both Majorana fermions and canonical bosons. A prototypical example of a spin-1/2 model with a $U(1)$ symmetry is the $XX$ model, which can be mapped to a model of hard-core bosons with hopping only. In this special case, no additional bosonic degrees of freedom are required in the fermionic dual. In the large $U/t$ limit, the bosons in the BH model become effectively hard-core, so that $\tilde{n}_j$ reduces to $\Pp_j^-$, taking values 0 or 1, for filling factor $\rho_0\leq 1$. At unit filling, subject to the constraint $\prod_{e\subset\partial p}(-1)^{n_{f,e}}=1$, the ground state at $t =0$ coincides with that of the Majorana fermion surface code \cite{vijay2015majorana}. In particular, the ground state exhibits a fourfold degeneracy on the torus. 

\section{Discussion}
\label{sec:dicussion}
In this work, we obtain the fermionic dual of the BH model in terms of fermionic composite operators through fermionic gauging, i.e., a generalized JW transformation. Supported by DMRG calculations, we show that the gapless phase in 1D is described by Luttinger liquid theory, or equivalently the compact boson CFT. The spectrum of the fermionic dual agrees exactly with that of the original BH model, up to the choice of boundary conditions, while the scaling dimensions of correlation functions of several microscopic operators are consistent with the CFT predictions. The correlation function $\langle B_i^{\dagger} A_j \rangle $ plays a crucial role at both fractional and integer fillings. Its oscillation wave vector, $k_F = \pi \rho_0 \mod \pi$, is fixed by the filling factor of the fermionic composites, providing a manifestation of a generalized Luttinger's theorem for a Bose-Fermi mixture. Our duality generalizes both the microscopic duality between the extended hard-core BH model and the spinless Fermi-Hubbard model and the low-energy duality between the sine-Gordon and Thirring models.  The construction generalizes to 2D and higher dimensions and extends the construction of Majorana fermion surface codes through fermionic gauging of  $\bbz_2$ symmetries.

In the original 1D BH model, the superfluid–Mott transition is of the BKT type, driven by the proliferation of phase slips, or vortices in spacetime. At commensurate filling $\rho_0=p/q$, the allowed cosine perturbation in the sine-Gordon theory has scaling dimension $\Delta(\cos 2q\phi)=q^2K$ and becomes relevant for $K<2/q^2$, with the BKT threshold at $K_c=2/q^2$. At unit filling, where $q=1$, this perturbation corresponds to the chirality-changing Umklapp term $\psi_L^\dagger\psi_R+\mathrm{h.c.}$ in the Thirring description. In the fermionic dual of the BH model, the leading continuum components of $A_j$ and $B_j$ involve the left- and right-moving fermionic fields $\psi_L$ and $\psi_R$. In the hard-core limit, $A_j$ reduces to $f_j$ within the low-energy subspace, and the dual model becomes a trivial filled-band insulator at unit filling. Finite onsite occupancy beyond the hard-core constraint is therefore essential for the gapless phase at unit filling in the absence of additional interactions. It also generates effective fermionic interactions that renormalize the Luttinger parameter $K$. At unit filling, the BKT transition is driven by chirality-changing Umklapp scattering between the emergent left- and right-moving fermionic components of $A_j$, described by $\psi_L^\dagger\psi_R+\mathrm{h.c.}\sim\cos(2\phi)$. In the gapless phase, both $A_j^2$ and $\tilde b_j$ have leading contributions proportional to the same charge-2 vertex operator, $V_{0,2}$, and consequently exhibit quasi-long-range pairing order. This common continuum identification explains their matching correlation exponents without requiring a nonzero condensate or a causal hierarchy between the two operators. At unit filling, $K>2$ throughout the gapless phase, so that $\Delta(A^2)=1/K<\Delta(A)=(K+1/K)/4$. Pair correlations therefore decay more slowly than the leading fermionic correlations, giving the dual Luttinger liquid a pairing-dominated, superconducting-like character. At fractional filling $0<\rho_0<1$, by contrast, the large-$U/t$ limit approaches $K=1$, where fermionic correlations decay more slowly than pair correlations and the conventional free-fermion description is recovered.

To compare with conventional bosonic gauging, we return to the boundary-condition-dependent correspondence between the gapless phase of the BH model and its fermionic dual in 1D. In fermionic gauging, a pair of Majorana fermions, or equivalently a complex fermion, is inserted on each edge, with the two Majorana fermions assigned to the two neighboring sites. Consequently, neighboring sites do not have a shared Majorana fermion. The resulting fermionic theory has a $U(1)$ symmetry that contains fermion parity $\bbz_2^F$ as a subgroup. This differs from bosonic gauging, in which an Ising gauge spin is shared by two neighboring sites, leading to a mixed anomaly between the quotient $U(1)$ symmetry and the dual $\bbz_2$ symmetry on a periodic chain. See Appendix~\ref{sec:bosonic_gauging} for further discussion on this point. When formulated on a torus $M$,  bosonic $\bbz_2$-gauging amounts to summing over boundary conditions classified by the group $H^1(M, \bbz_2)$ in the bosonic partition function. By contrast, the fermionic partition function has an intrinsic dependence on spin structures or fermionic boundary conditions, which form a set $S(M)$. Since $S(M)$ is a principal homogeneous space or torsor for the group $H^1(M, \bbz_2)$, there is a noncanonical bijection between the sets of bosonic boundary conditions and fermionic spin structures within each charge/parity sector. Once such an identification is chosen, the exact duality gives a corresponding sector-by-sector identification of their partition functions \cite{karch2019, ji2020top}. 

While we have primarily focused on the vanilla BH model, the same analysis can be extended to the general extended BH model and other non-standard BH models \cite{chandaRecent2025}. In Appendix \ref{sec:fermionic_gauging}, we show that the fermionic gauging of the quantum rotor and clock models follows a closely analogous construction. Tuning various parameters in the general bosonic models allows access to a variety of quantum phases.  Since several bosonic models are known to be exactly solvable \cite{cazalillaOne2011a}, so are their fermionic duals. Our construction also naturally extends to more general symmetries, such as $U(1)^n$, as well as non-Abelian groups such as $Spin(n)$. Furthermore, our fermionic dual motivates the study of Bose-Fermi mixtures in optical lattices and other strongly correlated systems with conversion couplings, whose parameter spaces contain our family of dual Hamiltonians as a subsector. In general, such systems conserve the combined charge $Q=2N_{\tilde b}+N_f$, while the larger $U_b(1)\times U_f(1)$ symmetry is restored in the absence of conversion terms. The extension of our construction to the anyon-Hubbard model, with its recent experimental realizations, is also intriguing. Finally, the generalization of Luttinger's theorem to Bose-Fermi mixtures and its connection to 't Hooft anomalies warrant further investigation. The distinct behavior of correlation functions at integer filling also calls for a more comprehensive theoretical understanding.

\textit{Note added}: Upon finalizing our manuscript, two papers \cite{dharanikota2026, luFermionic2026} appeared on arXiv studying fermionic gauging of 1D Villain models, which are closely related to the quantum rotor model presented in Appendix~\ref{sec:fermionic_gauging}. Our fermionic dual of the BH model, together with these constructions, suggests a possible strategy for circumventing the assumptions of the Nielsen-Ninomiya no-go theorem \cite{nielsennogo1981}: realizing canonical fermionic fields at low energies using noncanonical fermionic composite operators on the lattice.

\section*{Acknowledgments}
LS thanks Arkya Chatterjee, Marton Lajer, Dmitriy Pavshinkin, Alexei Tsvelik, and Meng Zeng for helpful discussions. This work was supported in part by Brookhaven National Laboratory (Contract No.~DE-SC0012704; LS), the U.S. Department of Energy, Office of Science, Basic Energy Sciences, Materials Sciences and Engineering Division (LS and IM),  and the Simons Foundation (Grant No. 669487; LS and AC).

\appendix

\section{Bosonic parity decomposition}
\label{sec:decomposition}
The Fock space ${\cal{F}}_b$ of a single bosonic mode with creation (annihilation) operator $b^{\dagger}$ ($b$) is spanned by states with different occupancy $n_b = b^{\dagger} b$. The $\bbz_2$ subgroup of the $U(1)$ symmetry generated by $n_b$ is associated with the boson parity $(-1)^{n_b}$ of the Fock space. We can define 
\be 
\tilde{b} = \frac{1}{\sqrt{2(n_b+1)}} b^{2}, \quad  \tilde{b}^{\dagger} = b^{\dagger 2} \frac{1}{\sqrt{2(n_b+1)}}. 
\ee 
Then  $[\tilde{b} , \tilde{b}^{\dagger} ] = 1$ and $\tilde{n}_b = \tilde{b}^{\dagger} \tilde{b} = n_b/2$ within the even sector. Thus the original Fock space ${\cal{F}}_b$ splits into two invariant sectors with even and odd parity:
$ 
{\cal{F}}_b = {\cal{F}}_{\text{even}} \oplus {\cal{F}}_{\text{odd}}$ with ${\cal{F}}_{\tilde{b}} = {\cal{F}}_{\text{even}}$. We can also write ${\cal{F}}_b = {\cal{F}}_{\text{even}} \otimes \bbc^2$, and define another parity decomposition
\be 
\begin{split}
b &=  \sqrt{2}\tilde{b}  \sigma^+   +  \sqrt{2 \tilde{n}_{b} +1} \sigma^- , \\
b^{\dagger} &=  \sqrt{2}\tilde{b}^{\dagger} \sigma^-  +  \sqrt{2 \tilde{n}_{b} +1} \sigma^+.
\end{split}
\label{eq:split}
\ee 
Here, Pauli operators $\sigma^{\pm}$ connect sectors of different parity.

In the $\bbz_2$-gauging procedure, we need to construct a disentangling unitary by using a unitary parity-switching operator that anticommutes with $(-1)^{n_b}$: 
\be 
\begin{split}
K  &= \sum_{k \in \bbz_{\ge 0}} |2k\rangle\langle 2k+1| +|2k+1\rangle\langle 2k|   \\
&= \frac{1}{\sqrt{n_b}} b^{\dagger}  P^+_b + 
  b \frac{1}{\sqrt{n_b}} P^-_b  \\
  &= \begin{cases}  \frac{1}{\sqrt{n_b}} b^{\dagger} = b^{\dagger}  \frac{1}{\sqrt{n_b +1 }}  &   \ \text{parity even,}  \\
  b \frac{1}{\sqrt{n_b}} =  \frac{1}{\sqrt{n_b+1}}  b  &  \ \text{parity odd},
 \end{cases}     
\end{split}
\ee 
where $P^{\pm}_b =[1 \pm (-1)^{n_b}]/2$. Using the decomposition in Eq.~(\ref{eq:split}), we can identify $K = \sigma^x$. $K$ acts on ${\cal{F}}_b$ as 
\be 
K n_b K^{-1}= n_b + (-1)^{n_b},
\ee 
and
\be 
\begin{split}
    K b K^{-1} &= b^{\dagger}P^+_b + \frac{1}{\sqrt{(n_b+1)(n_b+3)}} b^3 P^-_b, \\
     K b^{\dagger} K^{-1}& = b P^-_b + \frac{1}{\sqrt{(n_b-2)n_b}}   b^{\dagger 3}   P^+_b .
\end{split}
\ee  
Within the even sector,
\be 
\begin{split}
b^{\dagger} K &= \sqrt{2}\tilde{b}^{\dagger}, \quad K b^{\dagger} = \sqrt{2\tilde{n}_b +1}, \\
K b &= \sqrt{2} \tilde{b}, \quad b K   =  \sqrt{2\tilde{n}_b +1}.
\end{split}
\ee 
These identities are useful for simplifying the derivations in this work. 

\section{$su(1,1)$ polynomial current algebra}
\label{sec:loopalgebbra}

For a single bosonic mode, one can define
\be
K_+ = \frac{1}{2}b^{\dagger2},\quad
K_- = \frac{1}{2}b^2,\quad
K_0 = \frac{1}{2}\left(n_b+\frac{1}{2}\right).
\ee
These operators generate a non-compact Lie algebra $su(1,1)$:
\be
[K_0,K_\pm]=\pm K_\pm,\quad [K_-,K_+]=2K_0.
\ee
The same algebra can also be realized for multiple modes. For two modes, for example,
\be
K_+=b_1^\dagger b_2^\dagger,\quad
K_-=b_1b_2,\quad
K_0=\frac{1}{2}(n_{b_1}+n_{b_2}+1).
\ee

The free boson model at $U=0$ is integrable and, in the thermodynamic limit, possesses infinitely many conserved local charges. As we demonstrate below, at $\mu=0$ its conserved quadratic operators also realize an $su(1,1)$ polynomial current algebra. We first consider the integrable Hamiltonian
\be
H'=-\sum_j(b_j^\dagger b_{j+1}+b_{j+1}^\dagger b_j)
-\mu\sum_j b_j^\dagger b_j.
\ee
The local conserved charges are given by
\be
\begin{aligned}
Q_m^{(+)}&=\sum_j(b_j^\dagger b_{j+m}+b_{j+m}^\dagger b_j),\\
Q_m^{(-)}&=i\sum_j(b_j^\dagger b_{j+m}-b_{j+m}^\dagger b_j).
\end{aligned}
\ee
At finite system size, only finitely many of these charges are linearly independent.

In the main text, we define $Q_0=\sum_j b_j^\dagger b_j$ and $Q'_1=i\sum_j(b_j^\dagger b_{j+1}-b_{j+1}^\dagger b_j)$. For $\mu=0$, we can also construct other charges, such as
\be
\begin{aligned}
Q_1&=i\sum_j(b_{2j-1}^\dagger+b_{2j-1})(b_{2j}^\dagger-b_{2j})\\
&\quad-i\sum_j(b_{2j}^\dagger-b_{2j})(b_{2j+1}^\dagger+b_{2j+1}).
\end{aligned}
\ee
This expression simplifies after a rotation of the Hamiltonian to
\be
H=i\sum_j(b_j^\dagger b_{j+1}-b_jb_{j+1}^\dagger),
\ee
for which
\be
Q_1=\sum_j(b_j^\dagger-b_j)(b_{j+1}^\dagger-b_{j+1}).
\ee
Correspondingly, $Q_0=\sum_j b_j^\dagger b_j$ and $Q'_1=\sum_j(b_j^\dagger b_{j+1}+b_{j+1}^\dagger b_j)$. At low energies, $Q_0$ and $Q'_1$ flow to the particle (momentum) and vortex (winding) charges of the compact boson CFT in the superfluid phase at finite $U/t$. In particular, DMRG calculations verify that the local density associated with $Q'_1$ has scaling dimension $\Delta=1$ in the superfluid phase, independent of $U/t$. In contrast to the hard-core boson or free-fermion case \cite{chatterjee2025quantized}, the other charge $Q_1$ is not quantized. Its local density is dominated by the terms $b^2$ and $b^{\dagger2}$, and therefore has scaling dimension $1/K$, which depends on $U/t$.

Moreover, $Q_0$ and $Q_1$ do not commute and generate a larger algebra of conserved quadratic operators. To see this, we apply the Fourier transform so that
\be
H=i\sum_j(b_j^\dagger b_{j+1}-b_jb_{j+1}^\dagger)
=-2\sum_q\sin(q)n_q,
\ee
and
\be
Q_0=\sum_q b_q^\dagger b_q,\quad
Q_1=\sum_q\cos(q)
\left(b_q^\dagger b_{-q}^\dagger+b_qb_{-q}-2b_q^\dagger b_q\right).
\ee
Because the modes $q$ and $-q$ form a single two-mode sector, it is convenient to choose a set $\mathcal B_+$ containing one representative of each generic pair $\{q,-q\}$. For $q\in\mathcal B_+$, define
\be
\begin{aligned}
K_+(q)&=b_q^\dagger b_{-q}^\dagger,\\
K_-(q)&=b_qb_{-q},\\
K_0(q)&=\frac{1}{2}(n_{b,q}+n_{b,-q}+1).
\end{aligned}
\ee
These operators satisfy an independent $su(1,1)$ algebra for each generic momentum pair,
\be
\begin{aligned}
[K_0(q),K_\pm(q')]&=\pm\delta_{q,q'}K_\pm(q),\\
[K_-(q),K_+(q')]&=2\delta_{q,q'}K_0(q).
\end{aligned}
\ee
For $q=0$ (and $q=\pi$ when present), the generators instead take the single-mode form given above. For simplicity, we suppress these special sectors below.

In terms of the momentum pairs,
\be
Q_0=\sum_{q\in\mathcal B_+}(2K_0(q)-1),
\ee
up to the omitted special modes, while
\be
Q_1=2\sum_{q\in\mathcal B_+}\cos(q)
\left[K_+(q)+K_-(q)-2K_0(q)+1\right],
\ee
again up to the special-mode contributions. The factor of two arises because the original sum runs over both $q$ and $-q$. The scalar terms do not affect the commutator algebra.

To describe the algebra containing these charges, introduce the conserved quadratic operators
\be
T_a^{(m)}=\sum_{q\in\mathcal B_+}\cos^m(q)K_a(q),
\qquad a\in\{0,+,-\},\quad m\ge0.
\ee
Introducing the spectral variable $z_q=\cos(q)$, their commutation relations are
\be
\begin{aligned}
[T_0^{(m)},T_\pm^{(n)}]&=\pm T_\pm^{(m+n)},\\
[T_-^{(m)},T_+^{(n)}]&=2T_0^{(m+n)}.
\end{aligned}
\ee
Thus, in the thermodynamic limit, the full family of these operators realizes the polynomial current algebra
\be
su(1,1)\otimes\mathbb C[z].
\ee
Strictly speaking, this is not the full loop algebra, which would involve Laurent powers $z^m$ with $m\in\mathbb Z$.

The subalgebra generated by $Q_0$ and $Q_1$ is smaller. Modulo scalar operators,
\be
Q_0=2T_0^{(0)},\qquad
Q_1=2(T_+^{(1)}+T_-^{(1)}-2T_0^{(1)}).
\ee
Repeated commutators generate $T_0^{(0)}$ and all $T_a^{(m)}$ with $m\ge1$, but not $T_\pm^{(0)}$. Thus, in the thermodynamic limit, the generated subalgebra is
\be
\mathfrak g=
\operatorname{span}\{T_0^{(0)}\}
+\operatorname{span}\{T_a^{(m)}:a\in\{0,+,-\},\,m\ge1\}.
\ee
At finite lattice size, the operators $T_a^{(m)}$ are not all linearly independent because $z_q=\cos(q)$ takes only finitely many values. The full polynomial current algebra therefore reduces to a finite-dimensional quotient, while $Q_0,Q_1$ generate the image of $\mathfrak g$ in this quotient. The infinite-dimensional algebras described above emerge in the thermodynamic limit.

As in the Onsager construction \cite{onsager1944}, we can let $A_0=Q_0$ and $A_1=Q_1$ and ask whether they satisfy the Dolan-Grady relations \cite{dolan1982}. With the present normalization,
\be
[A_0,[A_0,[A_0,A_1]]]=4[A_0,A_1],
\ee
so the first relation has the Dolan-Grady form up to normalization. However,
\be
[A_1,[A_1,[A_1,A_0]]]=0,
\ee
while $[A_1,A_0]\neq0$. Thus, the canonical-boson generators considered here do not satisfy the Dolan-Grady relations. For hard-core bosons or free fermions, the operator algebra is different. The corresponding quadratic pseudospin generators obey an $su(2)$-type algebra rather than the bosonic $su(1,1)$ algebra, and in the hard-core/free-fermion realization of Ref.~\cite{chatterjee2025quantized}, the analogous generators satisfy both Dolan-Grady relations and generate an Onsager algebra. This should not, however, be regarded as a continuous reduction $su(1,1)\to su(2)$: imposing the hard-core constraint changes the underlying operator algebra and gives a distinct algebraic realization.

In the fermionic dual of the BH model, the corresponding charges are $Q_0=\sum_j A_j^\dagger A_j$ and $Q'_1=-i\sum_j(B_j^\dagger A_{j+1}-A_{j+1}^\dagger B_j)$. They flow to the vector charge and the axial charge at low energies, respectively. The $Q_1$ charge at $U=0$ is
\be
\begin{aligned}
Q_1&=i\sum_j(B_{2j-1}-B_{2j-1}^\dagger)
(A_{2j}^\dagger-A_{2j})\\
&\quad+i\sum_j(B_{2j}^\dagger+B_{2j})
(A_{2j+1}^\dagger+A_{2j+1}).
\end{aligned}
\ee
By the exact duality, the commutator algebra of these charges is isomorphic to the bosonic algebra described above. In particular, the dual images of the full family $T_a^{(m)}$ realize the same $su(1,1)$ polynomial current algebra in the thermodynamic limit and its corresponding quotient at finite size. The dual charges $Q_0,Q_1$ generate the subalgebra $\mathfrak g$, or its finite-size image.

\section{Large-$U$ perturbation theory}
\label{sec:largeU}
For $U \to \infty$, the low-energy subspace of the Hamiltonian in Eq.~(\ref{eq:dualBH}) satisfies $\tilde{n}_j = 0, 1$. In this limit, $A_j$ reduces to $f_j$ within the subspace, and the Hamiltonian reduces to that of a free fermion model. For large $U$ and filling $\rho_0 \le 1$, we can perform a degenerate perturbation theory, or equivalently a Schrieffer-Wolff transformation \cite{schriefferRelation1966}, to obtain the effective Hamiltonian. 

Let $P$ be the projection operator onto the low-energy subspace: $|0\rangle, |1\rangle$. The perturbation is 
$V =   t  \sum_j    (B_j^{\dagger}A_{j+1}  +h.c.) $.
To first order, $H^{(1)} =PVP = - t \sum_j (f_j^{\dagger} f_{j+1} + h.c.)$. The second order is obtained by the degenerate perturbation theory or a Schrieffer-Wolff transformation $H_{\text{eff}} = e^{S} H e^{-S}$ with 
\be 
S = \frac{1}{H_0 -E_0} QVP  - PVQ \frac{1}{H_0 -E_0}.
\ee 
Here, $Q = 1- P$, and $H_0$ is the Hubbard interaction with ground-state energy $E_0$. The second-order term in the  effective Hamiltonian is given by 
\be 
H^{(2)} = PVQ \frac{1}{E_0 - QH_0 Q} QVP.
\ee 
The lowest high-energy states have one doublon  $\tilde{n}_j =2$ with energy $U$. Thus
\be 
H^{(2)} =  -\frac{1}{U}PVQVP.
\ee 
For an adjacent occupied pair, the two virtual doublon-holon states contribute $-4t^2/U$. For correlated next-nearest-neighbor hopping, the contribution is $2t^2/U$. Collecting the above terms, we obtain an effective Hamiltonian for fermion $f^{\dagger}_j$
\be
\begin{split}
 H_{\text{eff}}  \approx & \ - t \sum_j (f_j^{\dagger} f_{j+1} + h.c.) - \frac{4t^2}{U} \sum_j n_{f, j} n_{f, j+1} \\
 & + \frac{2t^2}{U} \sum_j ( f_{j-1}^{\dagger} n_{f, j} f_{j+1} + h.c. ).   
\end{split}
\ee  
This effective Hamiltonian differs from the nearest-neighbor Fermi-Hubbard model in Eq.~(\ref{eq:extendedBH}) by the third term. Nevertheless, virtual states involving $\tilde{b}_j$ generate effective interactions among neighboring fermions. Under the inverse Schrieffer-Wolff transformation, the effective fermion $f_j$, expressed in the original representation, becomes
 
\begin{align}
 \tilde f_j&\equiv e^{-S}f_je^S\\
  &
\approx f_j-\frac{\sqrt{2}t}{U}
\left[
\tilde b_{j-1}f_{j-1}^\dagger
+\tilde b_j(f_{j-1}^\dagger-f_{j+1}^\dagger)
-\tilde b_{j+1}f_{j+1}^\dagger
\right].     \nonumber 
\end{align} 
Thus, in the original Hilbert space, the operator that creates the low-energy quasiparticle is no longer the bare fermion, but rather a dressed fermionic operator involving nearby bosonic and fermionic degrees of freedom.

\section{Fermionic gauging}
\label{sec:fermionic_gauging}

\subsection{$U(1)$: Quantum rotor model} 
We first discuss fermionic gauging of the 1D quantum rotor model. A quantum rotor is defined by the canonical conjugate variables $(\phi, L)$ that satisfy $[\phi, L] = i$. Here, the phase is compact $\phi \sim \phi +2\pi$ and the Hilbert space is spanned by the eigenstates of the angular momentum $\{|m\rangle:  m\in \bbz  \}$, where $L|m\rangle = m|m\rangle$. The canonical commutation relation leads to  
\be 
\begin{split}
& e^{i\alpha L} \phi e^{- i\alpha L} = \phi + \alpha, \quad e^{i\beta \phi} L e^{- i\beta \phi} = L - \beta, \\
& e^{i\alpha L} e^{i \beta \phi} = e^{i \alpha \beta} e^{i \beta \phi}e^{i\alpha L}. 
\end{split}
\ee  
The Hamiltonian of the quantum rotor model is given by 
\be 
H = - 2J\sum_j \cos (\phi_j -\phi_{j+1}) + U \sum_j L_j^2.
\ee  
To gauge the rotor model, we insert a pair of Majorana fermions as in the main text and impose the Gauss law $ (-1)^{L_j}  (-1)^{n_{f, j}} =1$, 
with $(-1)^{n_{f,j}} = i\gamma_{j} \gamma'_{j}$, $\gamma'_{j} = i(f_j - f_j^{\dagger})$, and $ \gamma_{j} = (f_{j} + f_{j}^{\dagger})$. 
The minimally coupled Hamiltonian is 
\be 
H' = -J \sum_j (e^{i \phi_j} \Xi_{j+1/2} e^{-i \phi_{j+1}}  +h.c.) + U \sum_j L_j^2,
\ee  
where $\Xi_{j+1/2} \equiv i \gamma'_{j} \gamma_{j+1}$. Applying the unitary transformation
\be
\tilde{U} = \prod_j(\Pp_{j}^+ +\Pp_{j}^- e^{-i \phi_j})
\ee 
with $\Pp_{j}^{\pm} = [1\pm (-1)^{n_{f, j}}]/2$, we obtain $(-1)^{L_j} =1$ and 
\be 
L_j \to   L_j + \Pp_{j}^- =  L_j+n_{f,j}
\ee 
and 
\be 
\begin{split}
e^{i\phi_j} &\Xi_{j+1/2}e^{- i\phi_{j+1}} \\
\to  & \Xi_{j+1/2} e^{i\phi_j} (\Pp_{j}^+e^{-i \phi_j} +\Pp_{j}^- e^{i \phi_j})\\
\times & (\Pp_{j+1}^+e^{-i \phi_{j+1}} +\Pp_{j+1}^- e^{i \phi_{j+1}})e^{- i\phi_{j+1}}    \\
=&   -(f_j e^{2i \phi_j}-f_j^{\dagger}) (f_{j+1}^{\dagger}e^{-2i \phi_{j+1}} + f_{j+1} ), 
\end{split}
\ee 
where we have used the relations $\Pp^- = n_f$ and $\Pp^+ = 1- n_f$ for each fermionic mode. 
We can rescale $\tilde{L}_j = L_j/2$ and $\tilde{\phi}_j = 2\phi_j$, so $\tilde{L}_j$ can take any   integer values. We also define 
\be 
A_j  = e^{-i \tilde{\phi}_j} f_j^{\dagger}  +f_j , \quad  B_j  =  e^{-i \tilde{\phi}_j} f_j^{\dagger} -f_j = A_j (-1)^{n_{f, j}},
\ee
where we have used the relations $ f(-1)^{n_f} = - f$
 and $f^{\dagger} (-1)^{n_f}= f^{\dagger}$.
Then the fermionic dual  can be written as 
\be 
H'' =  J \sum_j (B_j^{\dagger} A_{j+1} +h.c.) + U \sum_j (2\tilde{L}_j + n_{f, j})^2. 
\ee 

An alternative way to obtain the fermionic dual makes use of an analog of the decomposition in Eq.~(\ref{eq:split}):   
\be 
e^{-i \phi_j}  \equiv    e^{-i \tilde{\phi}_j } \sigma_j^+ +\sigma_j^-.
\label{eq:split2}
\ee 
Define $|\psi_{2m +s}\rangle \equiv |m, s\rangle$,  
where $s =0$ for $\mid\downarrow\rangle$ and $s =1$ for $\mid\uparrow\rangle$. Then $e^{-i \phi_j}|\psi_{m, j}\rangle = |\psi_{m-1, j}\rangle$
and $L_j= 2\tilde{L}_j + P_j^+$, with $P_{j}^{\pm} = (1\pm \sigma_{j}^z)/2$. A standard JW transformation leads to the fermionic dual. 

\subsection{$\bbz_n$: Clock model}
Even though we have focused on $U(1)$ symmetry so far, we can derive an analogous fermionic dual for the $\bbz_n$ clock model starting from the quantum rotor model. For integer $n \ge 2$, define 
\be 
Z = e^{i \frac{2\pi}{n} L},\quad X = e^{i  \phi},
\ee 
and identify $|m +n\rangle \sim |m\rangle$, where $L|m\rangle = m|m\rangle$. One can then verify that $Z$ and $X$ satisfy the Heisenberg algebra
\be 
Z^n =1, \quad X^n =1, \quad Z X = \omega XZ \quad \text{with}\ \omega = e^{2\pi i/n}.
\ee 
The $\bbz_n$ clock model is governed by the Hamiltonian 
\be 
\begin{split}
  H & = - 2J\sum_j \cos (\phi_j -\phi_{j+1}) - 2h\sum_j \cos \frac{2\pi}{n}L_j \\
  & = - {J}\sum_j (X_j X_{j+1}^{\dagger} + X_j^{\dagger} X_{j+1})- {h}\sum_j (Z_j + Z_j^{\dagger}),  
\end{split}
\ee 
which reduces to the quantum rotor model in the limit $n \to \infty$ with $ h = {U n^2}/{(2\pi)^2}$. Therefore, for even $n$, the fermionic dual obtained by gauging $\bbz_2 \subset \bbz_n$ can be read off directly from the result in the previous section. For even $n$,  define $\tilde{n} = n/2$ and
\be 
\tilde{Z}_j = e^{i \frac{2\pi}{\tilde{n}} \tilde{L}_j},\quad \tilde{X}_j = e^{i  \tilde{\phi}_j}.
\ee 
Then $\tilde{Z}$ and $\tilde{X}$ satisfy the $\bbz_{\tilde{n}}$ Heisenberg algebra, with the identification $|m +\tilde{n}\rangle \sim |m\rangle$. The relations derived in the previous section then reduce to
\be 
A_j  = \tilde{X}_j^{\dagger} f_j^{\dagger}  +f_j , \quad  B_j  =  \tilde{X}_j^{\dagger}  f_j^{\dagger} -f_j = A_j (-1)^{n_{f, j}},
\ee
and the dual Hamiltonian is 
\be 
H'' =  J \sum (B_j^{\dagger} A_{j+1} +h.c.) -h \sum_j (\tilde{Z}_j e^{i \frac{2\pi}{n} n_{f, j}} +h.c.). 
\label{eq:zn1}
\ee 

We can also use the alternative approach mentioned above.  Using the analog of the decomposition in Eq.~(\ref{eq:split2}) and $L_j= 2\tilde{L}_j + P_j^+$ with $P_{j}^{\pm} = (1\pm \sigma_{j}^z)/2$, we obtain 
\be 
\begin{split}
&Z_j = e^{i \frac{2\pi}{n} (2 \tilde{L}_j +  P_j^+) } =   \tilde{Z}_j  (P_j^- +\omega P_j^+)  ,  \\
& X_j = e^{i \tilde{\phi}_j } \sigma_j^- +\sigma_j^+ = \tilde{X}_j \sigma_j^- +\sigma_j^+.
\end{split}
\ee 
Here, we have used the identity
$ e^{i\alpha P^{\pm}}  = P^{\mp} + e^{i\alpha} P^{\pm}$. 
Substituting these decompositions into the Hamiltonian and applying fermionic gauging or the JW transformation yields
\be 
\begin{split}
H''= &J \sum_j \left[ (\tilde{X}_j f_j  -f_j^{\dagger}) (\tilde{X}_{j+1}^{\dagger} f_{j+1}^{\dagger}  + f_{j+1} ) +  h.c. \right] \\
& - h \sum_j \left [ \tilde{Z}_j  (\Pp_j^+ +\omega \Pp_j^-)  + h.c.\right], 
\end{split}
\ee  
where $\Pp_{j}^{\pm} = [1\pm (-1)^{n_{f,j}}]/2$.  This is identical to Eq.~(\ref{eq:zn1}).

As a concrete example, we can gauge $\bbz_2 \subset \bbz_4$ of the $\bbz_4$ clock model. The gauged Hamiltonian is 
\be 
\begin{split}
    H'=&-J\sum_j\left(X_j \Xi_{j+1/2}  X_{j+1}^\dagger + h.c.\right)- 
    h\sum_j \left(Z_j+Z_j^\dagger\right), 
    \end{split}
\ee
where $Z^4_j=X^4_j=1$,  $Z_jX_j=iX_jZ_j$, and  $\Xi_{j+1/2} \equiv i \gamma'_{j} \gamma_{j+1} $, with the Gauss law $Z_j^2 (-1)^{n_{f,j}}=1 $. 
Using the disentangling unitary  $\tilde{U} = \prod_j(\Pp_{j}^+ +\Pp_{j}^- X^{\dagger}_j) $, we obtain 
\be 
Z_j^2 (-1)^{n_{f,j}} \to Z_j^2 =1, \ \ 
Z_j \to Z_j (\Pp_{j}^+ +i \Pp_{j}^- ) = Z_j e^{i \frac{\pi}{2} \Pp^-_j}.
\ee 
and 
\be
\begin{split}
   &  X_j  \Xi_{j+1/2} X_{j+1}^{\dagger} \\
   & \to  \Xi_{j+1/2}     (\Pp_{j}^+  +\Pp_{j}^- X_j^2)(\Pp_{j+1}^+X^{\dagger 2}_{j+1} +\Pp_{j+1}^- )\\
    & = - (X_j^2 f_j  -f_j^{\dagger}) (X^{2}_{j+1} f_{j+1}^{\dagger}  + f_{j+1} ).  
\end{split}
\ee   
Finally, identifying $\tilde{Z}_j = Z_j$ and $\tilde{X}_j = X_j^2$, which act on a two-level Hilbert space, we recover the Hamiltonian  $H''$ above. Conversely, one can verify that gauging the fermion parity in this dual theory recovers the original $\bbz_4$ clock model.

\subsection{Majorana fermion surface codes}
\label{sec:Maj}
In this section, we discuss fermionic gauging of a 2D Ising spin system with a $\bbz_2$ global symmetry to obtain Majorana codes.

Majorana stabilizer codes can exhibit some advantages over their qubit counterparts, including one-step \cite{vijay2015majorana, mclauchlanFermionParityBased2022} or fewer-step measurements \cite{karzig2017, litinski2018}, as well as encoding schemes involving both bosonic and fermionic twists
\cite{mclauchlannew2024}. They may also provide more efficient approaches for simulating fermionic systems \cite{li2018, viyuela2019}. In this section, we show how to derive some of them using fermionic gauging. Since the Majorana stabilizer code on the square lattice has been discussed in Ref.~\cite{su$mathbbZ_2$2025a}, we focus on the honeycomb lattice.

\begin{figure}[tb]
    \centering
    \includegraphics[width=0.99\linewidth]{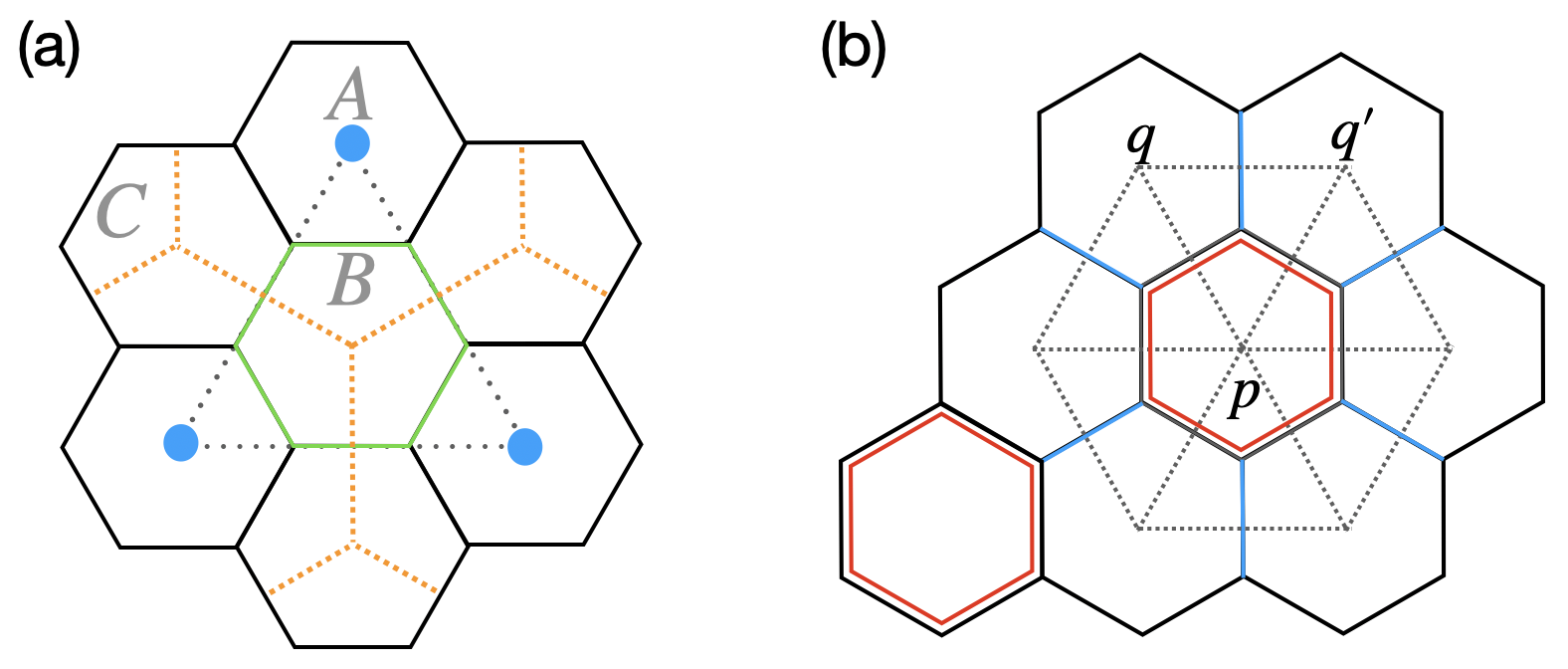}
    \caption{Equivalent representations of the Majorana code obtained by fermionically gauging the Ising Hamiltonian in Eq.~(\ref{eq:H0}). The original Ising spins are placed on the vertices of the dashed triangular lattice. (a) Majorana fermions on vertices. The hexagons are divided into three groups, $A$, $B$, and $C$. For $p\in A$, $O_p$ is associated with the $\sigma_p^x$, while for $p \in B, C$,  $O_p$ is derived from the flatness constraint. (b) Majorana fermions on edges. Each hexagon is surrounded by six red edges, with each edge representing a Majorana fermion. The blue edges correspond to the phase factors in Eq.~(\ref{eq:H1}). }
    \label{fig:gauging}
\end{figure}

Consider an Ising spin system with a $\bbz_2$ global symmetry on the triangular lattice. For concreteness, we take the Hamiltonian to be 
\be 
H_0 = - \sum_p \sigma_p^x.
\label{eq:H0}
\ee 
As shown in Fig.~\ref{fig:gauging}(a), the spins sit on vertices (blue dots) of the triangular lattice (dotted lines), or equivalently, on the faces of the dual honeycomb lattice (orange dashed lines). 
The Hamiltonian possesses a $\bbz_2$ symmetry generated by $\prod_p \sigma_p^x$.  In fermionic gauging, two Majorana fermions are placed on each edge, which are separated to form a honeycomb lattice (black lines), as represented by the endpoints of the green lines. Thus, each spin $\sigma^z_p$ on each (orange) dual hexagonal face $p$ is surrounded by six Majorana fermions $\gamma^e_p$. We impose the generalized Gauss law 
\be 
\sigma^x_p \left( i \prod_{e \subset \partial p} \gamma^e_p \right) =1,
\label{eq:gauging1}
\ee 
around each spin 
and a generalized flatness condition 
\be 
\prod_{\partial e \supset v} (-1)^{n_{f,e}} =1,
\label{eq:gausslaw200}
\ee 
on the six Majorana fermions surrounding each vertex of the orange honeycomb lattice, as represented by the green hexagon in Fig.~\ref{fig:gauging}(a). An ordering of Majorana fermions in the product is assumed. A disentangling unitary transformation as in Ref.~\cite{su$mathbbZ_2$2025a} can reduce the minimally coupled Hamiltonian to 
\be 
\tilde{H}_0 = - \sum_{p \in A} O_p, \quad O_{p} \equiv i \prod_{n \in \text{vertex}(p)} \gamma_n.
\label{eq:op}
\ee 
Here, we refer to the hexagon surrounding each spin as an $A$-type hexagon, while the remaining hexagons, exemplified by the green hexagon, are designated as $B$-type and $C$-type hexagons.
The flatness condition then takes the form
\be
i \prod_{n \in \text{vertex}(p)} \gamma_n =1 
\ee
for the six Majorana fermions on each $B$-type or $C$-type hexagon. Rather than enforcing this condition strictly, we can impose it softly by adding the corresponding terms to the Hamiltonian, giving
\be 
H'_0 = -\sum_p O_p,
\label{eq:H0p}
\ee 
up to coefficients. Note that all $O_p$ terms commute. The resulting
Hamiltonian is identical to that of the Majorana plaquette model studied in Ref.~\cite{vijay2015majorana}, which is a Majorana color-code model \cite{bravyiMajorana2010}.

The properties of the stabilizer code defined by Eq.~(\ref{eq:H0p}) have been discussed in detail in Ref.~\cite{vijay2015majorana}. Similar to the toric code, the ground state on the torus exhibits a fourfold degeneracy. There are three types of elementary plaquette excitations, corresponding to the three types of hexagons. On the torus, 
\be 
\Gamma = \prod_{p\in A} O_p = \prod_{p \in B} O_p = \prod_{p \in C} O_p
\ee 
is the total fermion parity. Consequently, conservation of total fermion parity implies that these three types of excitations can only be created or annihilated in pairs. The corresponding creation and annihilation operators are Wilson lines given by products of Majorana bilinears connecting plaquettes of the same type, such as $
W_A = \prod_{(mn)\in \beta_A}  i \gamma_m \gamma_n$, where $\beta_ A$ denotes a path connecting two $A$-type hexagons. By computing the braiding phases, one can show that the three types of elementary plaquette excitations have bosonic self-statistics and mutual semion statistics among them. These elementary excitations can also combine to form additional excitations, denoted by $AB$, $BC$, $AC$, and $ABC$. In particular, an  $ABC$  excitation is created by a $\gamma$ operator acting on a vertex and can therefore be interpreted as a physical fermion. By viewing $AB$, $BC$, $AC$ as composites of an elementary excitation with $ABC$, one can show that these excitations have fermionic self-statistics and mutual semion statistics with some elementary excitations. Moreover, the eight types of excitations can be divided into two groups, $(1, A, B, AB)$ and $ABC \times (1, A, B, AB)$, which are graded by fermion parity. The former set of anyons are equivalent to the anyon content of the toric code, a quantum double model, while the latter set is obtained by fusing these anyons with the physical fermion.  Together, these excitations form a super-modular tensor category.

There is an equivalent but more compact way to represent this model. We reduce the separation between the two Majorana fermions on the edge connecting two neighboring spins, and associate them with the orange (dual) edge in Fig.~\ref{fig:gauging}(a). The resulting effective lattice is then the one shown in Fig.~\ref{fig:gauging}(b). The $A$-type hexagons remain hexagons in Fig.~\ref{fig:gauging}(b), with each surrounded by six Majorana fermions residing on the red edges. The $B$-type and $C$-type hexagons now correspond to the three edges incident on the same vertex. Accordingly, the flatness condition is associated with the product of local fermion parities on the three edges. This dual lattice representation is more commonly encountered in conventional (bosonic) gauging \cite{su$mathbbZ_2$2025a}. 

Bosonic gauging can be used to distinguish different symmetry-protected topological (SPT) phases \cite{levin2012braiding}. This Levin-Gu diagnostic can likewise be generalized to fermionic gauging. For example, consider applying fermionic gauging to the topologically nontrivial Hamiltonian
\be 
H_1 =  \sum_p  \sigma_p^x \prod_{\langle p qq'\rangle} i^{\frac{1-\sigma^z_q \sigma_{q'}^z}{2}},
\label{eq:H1}
\ee 
where $p$ denotes a vertex of the dashed triangular lattice in Fig.~\ref{fig:gauging}(b), and the product runs over the six triangles $\langle pqq'\rangle$ containing $p$. The analog of Eq.~(\ref{eq:H0p}) after fermionic gauging in this case is a Majorana commuting-projector model. Under bosonic gauging, $H_1$ yields a topological order with four anyon types, $(1, A, B, AB)$, corresponding to the double-semion model.  Unlike the toric code, while $A$ remains bosonic, $B$ and $AB$ are semionic.  In the fermionic version, there are, in addition, physical fermionic excitations $ABC$, created by the action of $\gamma$ at a vertex. The second set of anyons is then given by $ABC \times (1, A, B, AB)$ obtained by fusing the anyons in the first set with the physical fermion. The corresponding braiding statistics can be verified using appropriately decorated string operators.

\section{Bosonic gauging}
\label{sec:bosonic_gauging}
For completeness, we also present the conventional bosonic gauging of the $\bbz_2$ subgroup for models discussed in this work. Compared with fermionic gauging, the distinguishing feature is the presence of mixed anomalies, which were studied in detail in Ref.~\cite{su2024}. In the main text of Ref.~\cite{su2024}, the Gauss law is imposed energetically to construct an intrinsically gapless SPT  phase. Consequently, such a disentangling transformation is not employed.  Instead, disentangling is discussed in Appendix B there using a slightly different approach. Here, we directly apply a disentangling unitary transformation to decouple the matter fields from the gauge spins in the Gauss law, providing a cleaner and more transparent derivation.

\subsection{$U(1)$: Quantum rotor model} 
Let us first gauge $\bbz_2 \subset U(1)$ of the 1D quantum rotor model
\be 
H = - 2J\sum_j \cos (\phi_j -\phi_{j+1}) + U \sum_j L_j^2,
\ee 
by inserting an Ising gauge spin on each edge and imposing the Gauss law $(-1)^{L_j} \sigma_{j-1/2}^z \sigma_{j+1/2}^z =1$. The minimally coupled Hamiltonian is 
\be 
H' = -J \sum_j (e^{i \phi_j} \sigma_{j+1/2}^x e^{-i \phi_{j+1}}  +h.c.) + U\sum_j L_j^2.
\ee  
Using the disentangling unitary 
\be 
\tilde{U} = \prod_j (\tilde{P}_{j}^+ + \tilde{P}_{j}^- e^{-i \phi_j} )
\ee 
with $\tilde{P}_{j}^{\pm} = (1\pm \sigma_{j-1/2}^z \sigma_{j+1/2}^z)/2$, we have the following mapping
\be 
(-1)^{L_{j}} \sigma_{j-1/2}^z \sigma_{j+1/2}^z \to (-1)^{L_{j}} =1, \quad L_j \to   L_j + \tilde{P}_{j}^-,
\ee  
\be 
\begin{split} 
&e^{i \phi_j} \sigma_{j+1/2}^x e^{-i \phi_{j+1}} \\
& \to   (\tilde{P}^+_{j} e^{2i\phi_j} +\tilde{P}^-_{j} ) \sigma_{j+1/2}^x (\tilde{P}^+_{j+1}  e^{-2i\phi_{j+1}} + \tilde{P}^-_{j+1} ) .  
\end{split} 
\ee 
After rescaling $\tilde{L}_j = L_j/2$ and $\tilde{\phi}_j = 2\phi_j$, the Hamiltonian becomes  
\be  
\begin{split}
  H'' = &   -J\sum_j [ (\tilde{P}^+_{j} e^{i\tilde{\phi}_j} +\tilde{P}^-_{j} ) \sigma_{j+1/2}^x\\
  \times & (\tilde{P}^+_{j+1}  e^{-i\tilde{\phi}_{j+1}} + \tilde{P}^-_{j+1} )+ h.c.] + U\sum_j (2\tilde{L}_j + \tilde{P}_{j}^-)^2.  
\end{split}
  \label{eq:rotor2}
\ee 
Note that $H''$ possesses a $\bbz_2$ symmetry generated by $\prod_j \sigma^x_{j+1/2}$ and a $\tilde{U}(1) = U(1)/\bbz_2$ symmetry generated by $\sum_j 2\tilde{L}_j + \tilde{P}_{j}^-$, as follows from $e^{i \pi(2\tilde{L}_j + \tilde{P}_{j}^-)} =\sigma_{j-1/2}^z \sigma_{j+1/2}^z$. On an open interval, \be 
U_{j_0\le j \le j_1}(\pi) =  \prod_{j_0\le j \le j_1} e^{i \pi(2\tilde{L}_j + \tilde{P}^-_{j})} = \sigma_{j_0-1/2}^z\sigma_{j_1+1/2}^z,
\ee   
i.e., the endpoints carry charges under the $\bbz_2$ symmetry generated by  $\prod_j \sigma^x_{j+1/2}$. This provides a manifestation of the mixed anomaly between $\bbz_2$ and $\tilde{U}(1) $, analogous to the mixed anomaly in the $\bbz_2$-gauged BH model discussed in later sections. 

We can also apply an alternative disentangling unitary to the minimally coupled Hamiltonian $H'$: 
\be 
\tilde{U}' = \prod_j (P_{j+1/2}^+ + P_{j+1/2}^- e^{i \phi_j}  e^{-i \phi_{j+1}} ),
\ee 
where $P_{j+1/2}^{\pm} = (1\pm \sigma_{j+1/2}^z)/2$. Then 
\be 
(-1)^{L_{j}} \sigma_{j-1/2}^z \sigma_{j+1/2}^z \to (-1)^{L_{j}} =1,
\ee 
\be 
L_j \to   L_j + P_{j-1/2}^- - P_{j+1/2}^-,
\ee
and 
\be  
\begin{split}
e^{i \phi_j} \sigma_{j+1/2}^x e^{-i \phi_{j+1}}  &\to   \sigma_{j+1/2}^x  (P^+_{j+1/2} e^{2i \phi_j - 2i\phi_{j+1}} + P^-_{j+1/2} )  \\
& = \sigma_{j+1/2}^x  e^{ 2i(\phi_j - \phi_{j+1}) P_{j+1/2}^+  }.  
\end{split}
\ee 
After rescaling $\tilde{L}_j = L_j/2$ and $\tilde{\phi}_j = 2\phi_j$, the Hamiltonian becomes 
\begin{widetext}
\be 
\begin{split}
  H'' = & - J \sum_j \sigma_{j+1/2}^x  (1 + P^+_{j+1/2} e^{i\tilde{\phi}_j - i\tilde{\phi}_{j+1}} + P^-_{j+1/2}  e^{-i \tilde{\phi}_j + i\tilde{\phi}_{j+1}} )  + U \sum_j (2\tilde{L}_j + P_{j-1/2}^- - P_{j+1/2}^-)^2  \\
  = &  -  J  \sum_j  \left[\sigma_{j+1/2}^x  \left( 1 +  \cos (\tilde{\phi}_j - \tilde{\phi}_{j+1})\right)  + \sigma^y_{j+1/2} \sin(\tilde{\phi}_j - \tilde{\phi}_{j+1}) \right]   + U\sum_j (2\tilde{L}_j + P_{j-1/2}^- - P_{j+1/2}^-)^2.
\end{split}
\label{eq:rotor3}
\ee 
\end{widetext}
Note that, in this form, the original $\bbz_2$ symmetry generator $\prod_j \sigma_{j+1/2}^x$ is mapped to the non-onsite operator $\prod_j \sigma_{j+1/2}^x  e^{ 2i(\phi_j - \phi_{j+1}) P_{j+1/2}^+  } $. Since $U_{j_0\le j\le j_1}(\pi) = \sigma_{j_0-1/2}^z\sigma_{j_1+1/2}^z $, the endpoints remain charged under the $\bbz_2$ symmetry.   

\subsection{$\bbz_n$: Clock model}
As in the rotor case, we can apply bosonic gauging to the $\bbz_n$-clock model with even $n$. Similar to fermionic gauging, we can obtain the dual Hamiltonian directly from Eq.~(\ref{eq:rotor2}): 
\be 
\begin{split}   
&H''  = \\
&- J \sum_j  \left[ (\tilde{P}_j^+ \tilde{X}_j + \tilde{P}_j^-)\sigma_{j+1/2}^x (\tilde{P}_{j+1}^+ \tilde{X}_{j+1}^{\dagger} + \tilde{P}_{j+1}^-) + h.c.\right] \\
&- h \sum_j \left[\tilde{Z}_j(\tilde{P}_{j}^+ +\omega \tilde{P}_{j}^- ) + h.c.\right].
\end{split}
\label{eq:zn2}
\ee 
where $\tilde{P}_{j}^{\pm} = (1\pm \sigma_{j-1/2}^z \sigma_{j+1/2}^z)/2$, $\tilde{Z}_j = e^{i \frac{2\pi}{\tilde{n}} \tilde{L}_j}$, and $\tilde{X}_j = e^{i  \tilde{\phi}_j} $. Here, $\omega = e^{2\pi i/n}$, $\tilde{n} = n/2$, and $\tilde{Z}$ and $\tilde{X}$ satisfy the $\bbz_{\tilde{n}}$ Heisenberg algebra with the identification $|m +\tilde{n}\rangle \sim |m\rangle$. 

For completeness, let us derive the Hamiltonian by gauging the $Z_n$-clock model directly. The minimally coupled Hamiltonian is  
\be 
\begin{split}
    H'=&-J\sum_j\left(X_j\sigma^x_{j+1/2} X_{j+1}^{\dagger} + h.c.\right)- h\sum_j \left(Z_j+Z_j^\dagger\right) 
    \end{split}
\ee
with the  Gauss law $Z_j^{\tilde{n}} \sigma_{j-1/2}^z \sigma_{j+1/2}^z=1$. Applying the disentangling unitary
\be
\tilde{U} = \prod_j(\tilde{P}_{j}^+ +\tilde{P}_{j}^- X_j^{\dagger}),
\ee 
with  $\tilde{P}_{j}^{\pm} = (1\pm \sigma_{j-1/2}^z\sigma_{j+1/2}^z)/2$, we obtain  
\be 
Z_j^{\tilde{n}} \sigma_{j-1/2}^z \sigma_{j+1/2}^z \to Z_j^{\tilde{n}} =1, \quad 
Z_j \to Z_j  (\tilde{P}_{j}^+ +\omega \tilde{P}_{j}^- )  
\ee 
\be 
X_j \sigma^x_{j+1/2} X_{j+1}^{\dagger} \to   (\tilde{P}_j^+ X_j^2 + \tilde{P}_j^-)\sigma_{j+1/2}^x (\tilde{P}_{j+1}^+ X_{j+1}^{\dagger 2} + \tilde{P}_{j+1}^-). 
\ee
After the substitutions $\tilde{Z}_j = Z_j$ and $\tilde{X}_j = X_j^2$, where $\tilde{Z}_j$ and $\tilde{X}_j$ act on a $\tilde{n}$-dimensional Hilbert space, we obtain the same Hamiltonian.

Alternatively, we use a different disentangling unitary
\be
\tilde{U}' = \prod_j(P_{j+1/2}^+ +P_{j+1/2}^- X^{\dagger}_j X_{j+1}),
\ee 
with $P_{j+1/2}^{\pm} = (1\pm \sigma_{j+1/2}^z)/2$. Then
\be 
 Z_j^{\tilde{n}} \sigma_{j-1/2}^z \sigma_{j+1/2}^z \to Z_j^{\tilde{n}} =1, 
\ee 
\be 
\begin{split}
Z_j & \to Z_j  (P_{j-1/2}^+ +\bar{\omega}  P_{j-1/2}^- ) (P_{j+1/2}^+ +\omega P_{j+1/2}^-) \\
& = Z_j e^{i \frac{2\pi}{n} (P_{j+1/2}^- -P_{j-1/2}^-) },  
\end{split}
\ee 
and
\be 
X_j  \sigma^x_{j+1/2} X_{j+1}^{\dagger}  \to  \sigma^x_{j+1/2}(P^+_{j+1/2} + P^-_{j+1/2} X_j^{2} X_{j+1}^{\dagger 2} ). 
\ee
After the replacement $\tilde{Z}_j = Z_j$ and $\tilde{X}_j = X_j^2$, the Hamiltonian becomes 
\be 
\begin{split}
& H'''  = \\
&- J \sum_j  \sigma^x_{j+1/2}(1+ P^-_{j+1/2} \tilde{X}_j \tilde{X}_{j+1}^{\dagger} + P^+_{j+1/2} \tilde{X}_j^{\dagger} \tilde{X}_{j+1}) \\
&- h \sum_j \left[\tilde{Z}_j e^{i \frac{2\pi}{n} (P_{j+1/2}^- -P_{j-1/2}^-) }  + h.c.\right],  
\end{split}
\label{eq:zn4}
\ee 
which is the clock-model counterpart of Eq.~(\ref{eq:rotor3}), with the opposite orientation of the alternative disentangling unitary. 

In the case $n =4$, the Hamiltonian simplifies slightly, since $\tilde{Z}_j = \tilde{Z}_j^{\dagger}$ and $\tilde{X}_j = \tilde{X}_j^{\dagger}$.  The Hamiltonian in Eq.~(\ref{eq:zn2}) becomes
\be 
\begin{split}
& H''  =\\
& - J \sum_j  \left[ (\tilde{P}_j^+ \tilde{X}_j + \tilde{P}_j^-)\sigma_{j+1/2}^x (\tilde{P}_{j+1}^+ \tilde{X}_{j+1} + \tilde{P}_{j+1}^-) + h.c.\right] \\
& - 2h \sum_j  \tilde{Z}_j \tilde{P}_{j}^+,   
\end{split}
\label{eq:rotor4}
\ee  
and Eq.~(\ref{eq:zn4}) becomes
\be  
\begin{split}
    H'''=& -J\sum_j\ \sigma^x_{j+1/2}(1+ \tilde{X}_j \tilde{X}_{j+1}) \\
    &- h\sum_j  \tilde{Z}_j (1 + \sigma_{j-1/2}^z \sigma_{j+1/2}^z).  
\end{split}
    \label{eq:rotor5}
\ee 
We can further gauge the quotient $\bbz_2=\bbz_4/\bbz_2$ symmetry generated by $\prod_j \tilde{Z}_j$ to obtain 
\be 
\begin{split} 
   H''''=& - J\sum_j\ \sigma^x_{j+1/2}(1+  \tau _{j+1/2}^x)  \\
    &- h\sum_j (1 + \sigma_{j-1/2}^z \sigma_{j+1/2}^z) \tau _{j-1/2}^z\tau _{j+1/2}^z  . 
\end{split}
\ee
Defining 
\be  
\begin{split} 
    X'_{j+1/2}  &\equiv \tau ^z_{j+1/2} (P^-_{j+1/2} -i P^+_{j+1/2}), \\
    Z'_{j+1/2}  &\equiv \sigma_{j+1/2}^+ + \tau _{j+1/2}^x \sigma_{j+1/2}^-,  
\end{split}
\ee 
which satisfy the $\bbz_4$ Heisenberg algebra $(X')^4=(Z')^4=1$ and $Z'X'=iX'Z'$, 
we see that the model is precisely the dual $\bbz_4$ clock model. Note that the two Hamiltonians in Eq.~(\ref{eq:rotor4}) and Eq.~(\ref{eq:rotor5}) are related by a unitary transformation. The second form, however, is particularly appealing because the duality is manifest: 
\be \sigma^x_{j+1/2} \leftrightarrow \tilde{Z}_j,\quad \sigma^z_{j+1/2} \leftrightarrow \tilde{X}_j, \quad J \leftrightarrow h.
\ee 
In particular, the model is self-dual at the critical point $J =h$ and possesses a manifest noninvertible symmetry \cite{thorngren2024fusion}. Under this symmetry, the two $\bbz_2$ symmetries generated by $\prod_j \sigma_{j+1/2}^x$ and $\prod_j \tilde{Z}_j$ are switched. Since both symmetry generators are onsite, there is no mixed anomaly between them. One may wonder where the mixed anomaly between the quotient $\bbz_2$ symmetry and the dual $\bbz_2$ symmetry is hidden. Notice that under the unitary transformation, the original dual symmetry generator $\prod_j \sigma_{j+1/2}^x$ is mapped to 
\be
\begin{split}
    &\prod_j \sigma^x_{j+1/2}(P^+_{j+1/2} + P^-_{j+1/2} \tilde{X}_j \tilde{X}_{j+1}) \\
    =& \prod_j \sigma^x_{j+1/2} \exp[-i \pi P^-_{j+1/2} \tilde{P}_{j+1/2}^{X -}]
\end{split}
\ee with $\tilde{P}_{j+1/2}^{X -} = (1 -\tilde{X}_j \tilde{X}_{j+1})/2$. This non-onsite symmetry has a mixed anomaly with the $\bbz_2$ symmetry generated by $\prod_j \tilde{Z}_j$, making the critical point a deconfined quantum critical point \cite{zhang2023exactly, suBoundary2023a}. The mixed anomaly is encoded in the factor $\tilde{P}_{j+1/2}^{X -}$,  which detects decorated domain walls  \cite{li2024decorated}. If $n \ge 6$ with  $\tilde{n} = n/2$ even, tuning $h/J$ instead gives rise to a finite critical regime in which neither the quotient nor the dual symmetry is spontaneously broken. The resulting deconfined critical region corresponds to an intrinsically gapless SPT phase (see Appendix A in Ref.~\cite{su$mathbbZ_2$2025a}).

\subsection{$U(1)$: Bose-Hubbard model (1D)}
In this section, we apply $\bbz_2$-bosonic gauging to the 1D BH model 
\be 
H = -t \sum_j (b^{\dagger}_j b_{j+1} +h.c.) + \frac{U}{2} \sum_j n_{b,j}(n_{b,j} -1).
\ee 
Impose the Gauss law $(-1)^{n_{b, j}} \sigma_{j-1/2}^z \sigma_{j+1/2}^z =1 $ and write down the minimally coupled Hamiltonian
\be 
H' = -t \sum_j (b^{\dagger}_j \sigma_{j+1/2}^x b_{j+1} +h.c.) + \frac{U}{2} \sum_j n_{b,j}(n_{b,j} -1).
\ee    
Apply the disentangling unitary 
\be 
\tilde{U} = \prod_j (P_{j+1/2}^+ + P_{j+1/2}^- K_j K_{j+1})
\ee 
with $P_{j+1/2}^{\pm} = (1\pm \sigma_{j+1/2}^z)/2$, or \be 
\tilde{U}' = \prod_j (\tilde{P}_{j}^+ + \tilde{P}_{j}^- K_j )
\ee 
with  
$\tilde{P}^{\pm}_{j} = (1 \pm \sigma_{j-1/2}^z\sigma_{j+1/2}^z)/{2}$, then 
\be 
(-1)^{n_{b, j}} \sigma_{j-1/2}^z \sigma_{j+1/2}^z \to (-1)^{n_{b, j}} =1,  
\ee 
\be 
 n_{b, j}  \to  n_{b, j} + \tilde{P}^-_{j},  
 \ee 
\be 
\sigma_{j+1/2}^x \to \sigma_{j+1/2}^x K_j K_{j+1}, \quad  b^{\dagger}_j \to   \tilde{P}^{+}_{j}b^{\dagger}_j  +\tilde{P}^{-}_{j}K_j b^{\dagger}_j K_j.
\ee  
As a result, the Hamiltonian becomes
\be H'' =  -t \sum_j (A_j^{\dagger} \sigma_{j+1/2}^x  A_{j+1} +h.c.)  + \frac{U}{2} \sum_j \tilde{n}_j(\tilde{n}_j   -1),
\ee  
where  $A_j =   (\sqrt{2}\tilde{b}_{j} \tilde{P}_{j}^+  +    \sqrt{2\tilde{n}_{b, j} +1}\tilde{P}_{j}^- )$ and  $\tilde{n}_j = A^{\dagger}_j A_j = 2\tilde{n}_{b, j} + \tilde{P}^-_{j}$. We can obtain the same Hamiltonian by using the parity decomposition   $b_j \equiv (\sqrt{2}\tilde{b}_{j} \sigma^+_{j}  +  \sqrt{2 \tilde{n}_{b, j} +1} \sigma^-_{j}) $ in the BH Hamiltonian and gauging the $\bbz_2$ symmetry. 
Since 
$(-1)^{\tilde{P}^-_{j}} = \sigma_{j-1/2}^z\sigma_{j+1/2}^z $, the Hamiltonian possesses a $\tilde{U}(1) = U(1)/\bbz_2$ symmetry with    
$U(\pi) = \prod_j e^{i \pi (2\tilde{n}_{b, j} + \tilde{P}^-_{j})} =1$, which on an interval becomes
\be 
U_{j_0\le j\le j_1}(\pi) =  \prod_{j_0\le j\le j_1} e^{i \pi(2\tilde{n}_{b, j} + \tilde{P}^-_{j})} = \sigma_{j_0-1/2}^z\sigma_{j_1+1/2}^z.
\ee 
Both endpoints are charged under the dual $\bbz_2$ symmetry generated by $\prod_j \sigma_{j+1/2}^x$. In fact, the endpoints of $U_{j_0\le j \le j_1}(\alpha)$ carry effective fractional charge, providing a manifestation of the mixed anomaly between $\tilde{U}(1)$ and $\bbz_2$. In particular, gauging either symmetry necessarily breaks the other. Another manifestation of this anomaly is that two symmetries cannot be realized onsite simultaneously. As a consequence of the mixed anomaly, the ground state cannot be trivially gapped. Thus, in a gapped phase,  the $\bbz_2$ symmetry must be spontaneously broken, since $\tilde{U}(1)$ cannot be spontaneously broken in 1D due to the Mermin-Wagner theorem. Similar to the fermionic case discussed in the main text, the gapless phase is also described by the compact boson CFT \cite{su2024}.
Note that the mixed anomaly also implies that the boundary must be degenerate due to symmetry breaking when symmetries are not explicitly broken \cite{su2024}. This is because it is impossible to construct a symmetry-preserving boundary state for an anomalous theory \cite{hanBoundary2017a}. The superfluid phase can be viewed as an intrinsically gapless SPT phase if we regard the Gauss law as an emergent constraint \cite{su2024}. From this perspective, the mixed anomaly can be understood as an emergent anomaly of $\tilde{U}(1) \times \bbz_2$. Since the microscopic bulk theory is anomaly-free, the boundary theory must  cancel the emergent anomaly in the bulk to be consistent \cite{thorngrenIntrinsically2021, su2024}.

\subsection{$U(1)$: Bose-Hubbard model (2D)}
Let us generalize the discussion to the 2D BH model on the square lattice, whose Hamiltonian is given by
 \be 
H = -t \sum_{\langle i, j\rangle } ( b^{\dagger}_i b_{j} + b_i b_{j}^{\dagger})   + \frac{U}{2} \sum_j n_j  (n_j -1).  
\ee
Impose the  Gauss law 
$ 
(-1)^{n_{b, j}} \prod_{\partial e \supset j} \sigma_e^z  =1
$ 
on the minimally coupled Hamiltonian
\be 
H' = -t \sum_{\langle i, j\rangle }  (b^{\dagger}_i \sigma_{e_{ij}}^x b_{j} +h.c.) + \frac{U}{2} \sum_j n_{b,j}(n_{b,j} -1),
\ee    
with a flatness condition $\prod_{e \subset \partial p} \sigma_e^x =1$ for each plaquette $p$. Now apply the disentangling unitary 
\be 
\tilde{U} = \prod_j (\tilde{P}_j^+ + \tilde{P}_j^- K_j)
\ee 
with $\tilde{P}_j^{\pm} = (1\pm \prod_{\partial e \supset j} \sigma_e^z)/2$. The procedure yields the Hamiltonian 
\be 
H'' = -t \sum_{\langle i, j\rangle } (    A_i^{\dagger} \sigma_{e_{ij}}^x A_{j} +h.c.) + \frac{U}{2} \sum_j \tilde{n}_{j}(\tilde{n}_{j} -1).
\ee  
where $A_j = ( \sqrt{2}\tilde{b}_j \tilde{P}_j^+    +  \sqrt{2\tilde{n}_{b, j} +1}  \tilde{P}_j^- )$ and $\tilde{n}_{j} = A_j^{\dagger} A_j = 2 \tilde{n}_{b, j} + \tilde{P}_j^-$. The flatness condition $\prod_{e \subset \partial p} \sigma_e^x =1$ now is interpreted as the new Gauss law.   The dual Hamiltonian has  a dual  $\bbz_2^{(1)}$ 1-form symmetry generated by $\prod_{e \subset \gamma} \sigma^x_{e}$ along loops $\gamma$ on the square lattice and a quotient $\tilde{U}(1) = U(1)/\bbz_2$ symmetry generated by $\sum_j\tilde{n}_j$. The mixed anomaly between them leads to nontrivial consequences as in 1D \cite{su2024}.

\section{Edge-to-edge correlations}
\label{sec:edgetoedge}
\begin{figure}[tb]
    \centering
    \includegraphics[width=0.99\linewidth]{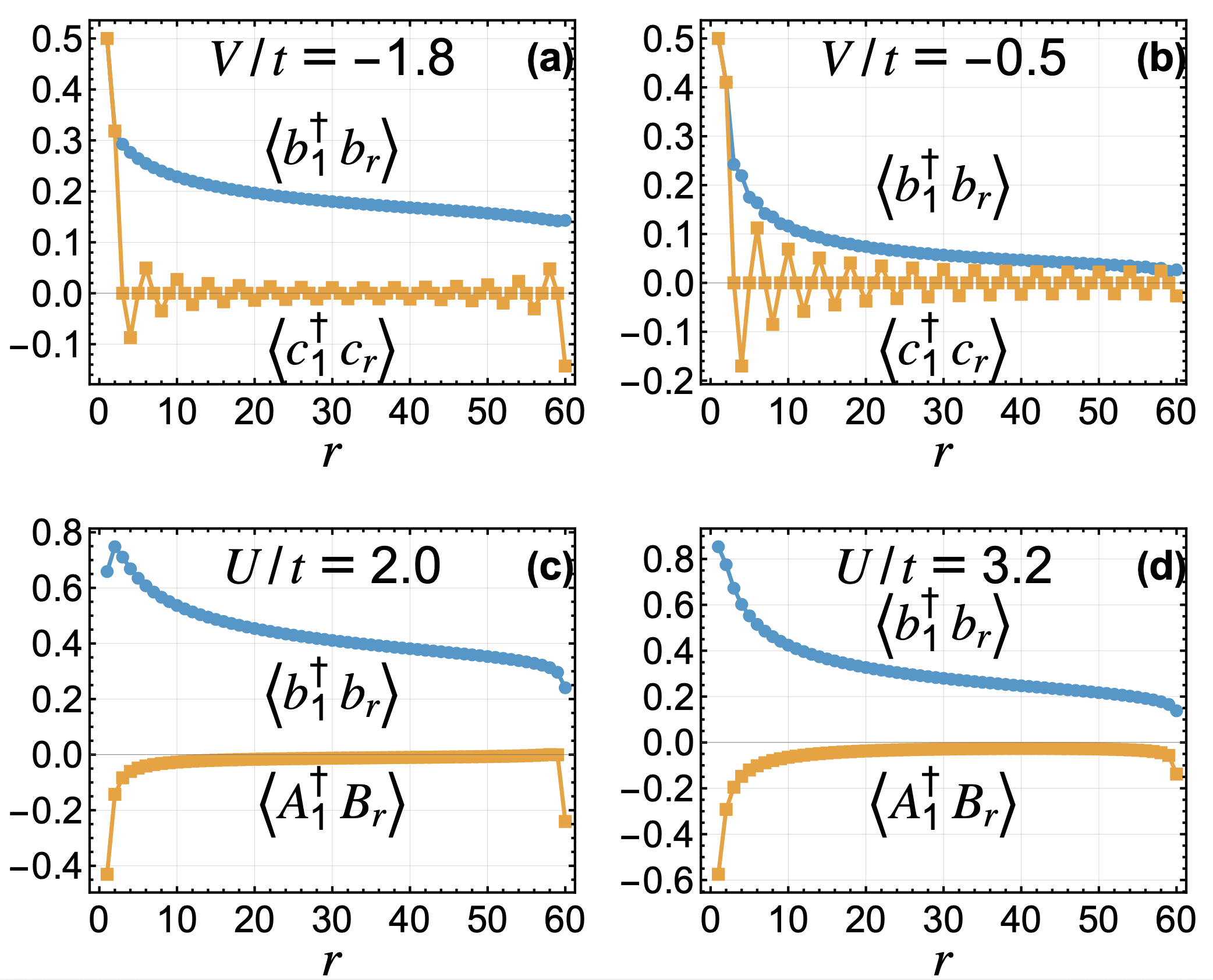}
    \caption{Edge-to-edge correlations in BH models and their fermionic duals under OBCs.  (a, b) $\langle b_1^{\dagger} b_r\rangle$ in the extended hard-core BH model (blue) and $\langle c_1^{\dagger} c_r\rangle$ in the fermionic dual (orange) at half filling and different $V/t$. The critical value is $V/t =-2$, where the scaling dimension of $b$ vanishes. (c, d)  $\langle b_1^{\dagger} b_r\rangle$ in the canonical BH model (blue) and $\langle A_1^{\dagger} B_r\rangle$ in the fermionic dual (orange) at unit filling and different $U/t$. In the limit of $U/t \to 0$, the scaling dimension of $b$ vanishes.  A noticeable drop in  $\langle b_1^{\dagger} b_r\rangle$ around $r \sim L$ is due to depletion in boson density $\langle b_r^{\dagger} b_r\rangle$ near the edges.}
    \label{fig5}
\end{figure}

When computing the correlation functions under OBCs to extract critical exponents, we use an interval centered symmetrically within the chain. In the main text, we mentioned that there are revivals in the edge-to-edge correlation function $\langle B^{\dagger}_1 A_L\rangle$ and claimed that these revivals arise from boundary effects and vanish in the thermodynamic limit. Here, we provide more details.

We first discuss the edge-to-edge correlations in the extended hard-core BH model and its fermionic dual shown in Eq.~(\ref{eq:extendedBH}). A phase transition occurs at $|V/t| =2$ at half filling. Close to the critical value $V/t =-2$, a noticeable revival in $\langle c_1^{\dagger} c_r\rangle$ appears around $r =L$ even for large system sizes [see Fig.~\ref{fig5}(a)], suggesting the possibility of the existence of an edge mode. However, the feature gradually dies out when $V/t$ moves away from the critical point [Fig.~\ref{fig5}(b)]. Mapping to the hard-core BH model, we can understand the cause of such a revival more easily as $\langle c_1^{\dagger} c_L\rangle$ is equal to $\langle b_1^{\dagger} b_L\rangle$ up to a sign determined by the total number parity. When $V/t$ is close to the critical point, the bosonic edge-to-edge correlation function decays very slowly as the scaling dimension approaches zero [see Fig.~\ref{fig5}(a)]. The fermionic counterpart decays very fast but is forced to match the value of the $\langle b_1^{\dagger} b_L\rangle$ at the other end because the JW string becomes rigid, leading to a revival of $\langle c_1^{\dagger} c_r\rangle$ as $r \to L$. As $V/t$ is tuned far away from the critical value [see Fig.~\ref{fig5}(b)], $\langle b_1^{\dagger} b_r\rangle$ decays faster and the revival of $\langle c_1^{\dagger} c_r\rangle$ is less prominent.  Since the bosonic correlation function $\langle b_1^{\dagger} b_r\rangle$ has a power-law decay and thus vanishes in the $L \to \infty$ limit for $-2 <V/t <2$, we conclude that $\langle c_1^{\dagger} c_L\rangle$ should reduce to zero eventually in the thermodynamic limit.

The above discussion generalizes to the soft-core BH model and its fermionic dual. The edge-to-edge correlation function satisfies 
$\langle b^{\dagger}_1b_L\rangle \sim \langle A_1^{\dagger} \prod_{1\le k < L} (-1)^{\tilde{n}_k} A_L\rangle \sim \prod_{1\le k \le L} (-1)^{\tilde{n}_k} \langle A_1^{\dagger} B_L\rangle $. Since the particle number parity is fixed, revivals in  $\langle A_1^{\dagger} B_L\rangle$ (or $\langle  B_1^{\dagger} A_L\rangle$) are associated with the decay of the correlation function $\langle b^{\dagger}_1b_L\rangle$. In Fig.~\ref{fig5}(c, d), we show the results at unit filling. The local density $\langle b_r^{\dagger} b_r\rangle$ exhibits boundary-induced Friedel-like oscillations and deviates from the bulk value near the edges, as reflected in the reduced value of $\langle b_1^{\dagger} b_1\rangle \approx 0.66 <1$. This boundary depletion leads to a noticeable suppression of $\langle b_1^{\dagger} b_L\rangle$. The revivals in $\langle A_1^{\dagger} B_L\rangle$ are sharp and  persist even for very large system sizes when $U/t$ approaches zero, where the scaling dimension of $b$ becomes very small. However, these revivals should not be interpreted as evidence for stable edge modes. Since the bosonic correlation function decays algebraically to zero, the relation above implies that $\langle A_1^{\dagger} B_L\rangle \to 0$ when $L \to \infty$.

\bibliography{refs}

\end{document}